\documentclass[twocolumn,aps,prb,floatfix,superscriptaddress]{revtex4}
\usepackage{amssymb}
\usepackage{color}
\usepackage{graphicx}
\usepackage{dcolumn}
\usepackage{bm}
\usepackage{epstopdf}
\usepackage[header,title,page,titletoc]{appendix}
\usepackage[colorlinks=true, letterpaper=true, pdfstartview=FitV, linkcolor=red, citecolor=blue, urlcolor=blue]{hyperref}
\usepackage{amsmath}
\usepackage{amsfonts}%
\usepackage{epstopdf}
\providecommand{\U}[1]{\protect \rule{.1in}{.1in}}
\begin{document}

\title{Accumulation of scale-free localized states induced by local non-Hermiticity}
\author{Cui-Xian Guo}
\affiliation{Beijing National Laboratory for Condensed Matter Physics,
Institute of Physics, Chinese Academy of Sciences, Beijing 100190, China}

\author{Xueliang Wang}
       \affiliation{Beijing National Laboratory for Condensed Matter Physics, Institute of Physics, Chinese Academy of Sciences, Beijing 100190, China}
\affiliation{School of Physical Sciences, University of Chinese Academy of Sciences, Beijing 100049, China}

\author{Haiping Hu}
\email{hhu@iphy.ac.cn}
\affiliation{Beijing National Laboratory for Condensed Matter Physics, Institute of Physics, Chinese Academy of Sciences, Beijing 100190, China}
\affiliation{School of Physical Sciences, University of Chinese Academy of Sciences, Beijing 100049, China}

\author{Shu Chen}
\email{schen@iphy.ac.cn}
\affiliation{Beijing National Laboratory for Condensed Matter Physics,
Institute of Physics, Chinese Academy of Sciences, Beijing 100190, China}
\affiliation{School of Physical Sciences, University of Chinese Academy of Sciences, Beijing 100049, China}
%\affiliation{Yangtze River Delta Physics Research Center, Liyang, Jiangsu 213300, China}

\begin{abstract}
The bulk states of Hermitian systems are believed insensitive to local Hermitian impurities or perturbations except for a few impurity-induced bound states. Thus, it is important to ask whether \textit{local} non-Hermiticity can cause drastic changes to the original Hermitian systems. Here we address this issue affirmatively and present exact solutions for the double chain model with local non-Hermitian terms possessing parity-time ($\mathcal{PT}$) symmetry. Induced by the non-Hermiticity, the system undergoes a sequence of $\mathcal{PT}$-symmetry breakings, after which the eigenenergies appear in complex conjugate pairs. The associated extended bulk states then become scale-free localized and unidirectionally accumulated around the impurity. There exist mobility edges separating the residual extended states until a full scale-free localization of all eigenstates. Further increasing the non-Hermitity counter-intuitively brings the system to a $\mathcal{PT}$-restoration regime with fully real spectra except for a pair of complex bound states. We demonstrate that the local non-Hermiticity generated scale-free localization is a general phenomenon and can even survive the quasiperiodic disorder. Our results indicate that the bulk properties of the original Hermitian system can be globally reshaped by local non-Hermiticity.
\end{abstract}
\maketitle

\section{Introduction}
An essential subject of band theory is the study of the sensitivity of the energy spectrum and eigenstate to local perturbations, like impurities or defects and various boundary conditions. Generally speaking, a local impurity or domain wall would only induce a few bound states for Hermitian systems. The bulk energy spectra are insensitive to such local perturbations, with the eigenstates' localization properties staying unchanged. In topological phases of matter, nontrivial in-gap modes residing at the impurities/defects or system boundaries may appear, governed by certain bulk topological invariants. However, such an intuitive physical picture breaks down for some non-Hermitian systems. As a paradigm, the non-Hermitian skin effect (NHSE), i.e., the extreme sensitivity of energy spectra and eigenstates to the change of boundary conditions, has attracted intensive studies in the past few years \cite{SYao1,TELee,Yokomizo,Okuma,KZhang,LeeCH,Slager,GuoCX,Kunst,HJiang,LJin,Kou,Longhi-PRR,ZSYang,GongJB,YFYi,Alvarez,CSE,Imura,LiuCH,Yao2,HN,Gong}. Without any Hermitian counterparts, it is featured by the entirely distinct energy spectra under different boundary conditions and the condensation of eigenstates at system boundaries \cite{SYao1,TELee,Yokomizo,Okuma,KZhang,LeeCH,Slager,GuoCX,Kunst,HJiang,LJin,Kou,Longhi-PRR,ZSYang,GongJB,YFYi,Alvarez,CSE,Imura,LiuCH,Yao2,HN,Gong}, domain walls \cite{WYi,Rafi,ExpOptXue1} or impurities \cite{YXLiu,Roccati,LiuYX2020}.

The NHSE necessities the extension of band theory to its non-Bloch form by introducing the so-called generalized Brillouin zone \cite{SYao1,Yokomizo,Okuma,KZhang}. In this context, most previous studies focused on non-Hermitian systems with either global (i.e., the non-Hermitian terms have support on the whole lattice) nonreciprocal hoppings or gain/loss. A few exceptions include studies on the dynamical properties of quantum systems dissipatively coupled to baths at the boundary \cite{local1} or subject to local loss \cite{local2,local3,local4,local5,local6,local7,local8}. Two fundamental and interesting questions naturally arise: (1) Is there any paradigmatic and universal phenomenon akin to the NHSE emerge from local non-Hermiticity? (2) can a local non-Hermitian term (i.e., has support only on a few lattice sites) cause dramatic changes to the energy spectra and eigenstates for an otherwise Hermitian system? Addressing these issues would bridge a comprehensive understanding of both the Bloch and non-Bloch band theory and is also experimentally relevant thanks to the feasibility of local manipulations of non-Hermiticity (e.g., non-reciprocity and gain/loss) in various classical and quantum simulation platforms.

In this work, we give affirmative answers to these questions by analytically solving a $\mathcal{PT}$-symmetric double chain model with a local gain/loss term (It equally describes a Su-Schrieffer-Heeger (SSH) lattice with a single asymmetrical hopping). We show that increasing the strength of gain/loss ($\gamma$) drives the system from the $\mathcal{PT}$-unbroken regime with entirely real spectra into a $\mathcal{PT}$-broken regime with the appearance of paired complex-conjugated eigenenergies. The $\mathcal{PT}$-transition is through a sequence of exceptional points accompanied by the formation of scale-free localized (SFL) eigenstates. These SFL states, unidirectionally accumulated near the impurity, have localization length of the order of system size \cite{LiLH2021}. Separated by mobility edges, the SFL states and residual extended states coexist until a full scale-free localization of all eigenstates occurs. Further increasing $\gamma$, the complex eigenenergies gradually coalesce into real eigenenergies, with their associated eigenstates changing from SFL to extended. Last, the system enters into a $\mathcal{PT}$-restoration regime with entirely real spectra except for a pair of complex bound states.

We demonstrate that the local non-Hermiticity-induced scale-free localization is a general phenomenon, regardless of the specific models, the underlying $\mathcal{PT}$-symmetry, the coalescence of eigenstates or even a priori Bloch-band description of the underlying Hermitian systems,  as verified in the quasiperiodic Aubry-Andr\'{e} (AA) model and a single-impurity chain. We note the key differences between the SFL states and the non-Hermitian skin modes. With its localization length proportional to system size, the emergence of SFL states only requires local non-Hermiticity and goes beyond both Bloch and non-Bloch band descriptions.

The rest of the paper is organized as follows. In section II,  we demonstrate in details how scale-free localization is induced by local non-Hermiticity by exactly solving the  double chain model and SSH model with local non-Hermiticity. In section III, we unveil the generality of SFL states by studying the model of single-impurity chain with an imaginary on-site potential and the quasiperiodic AA model with local non-Hermiticity. Conclusions and discussions are given in the last section.
\begin{figure}[b]
\includegraphics[width=0.47\textwidth]{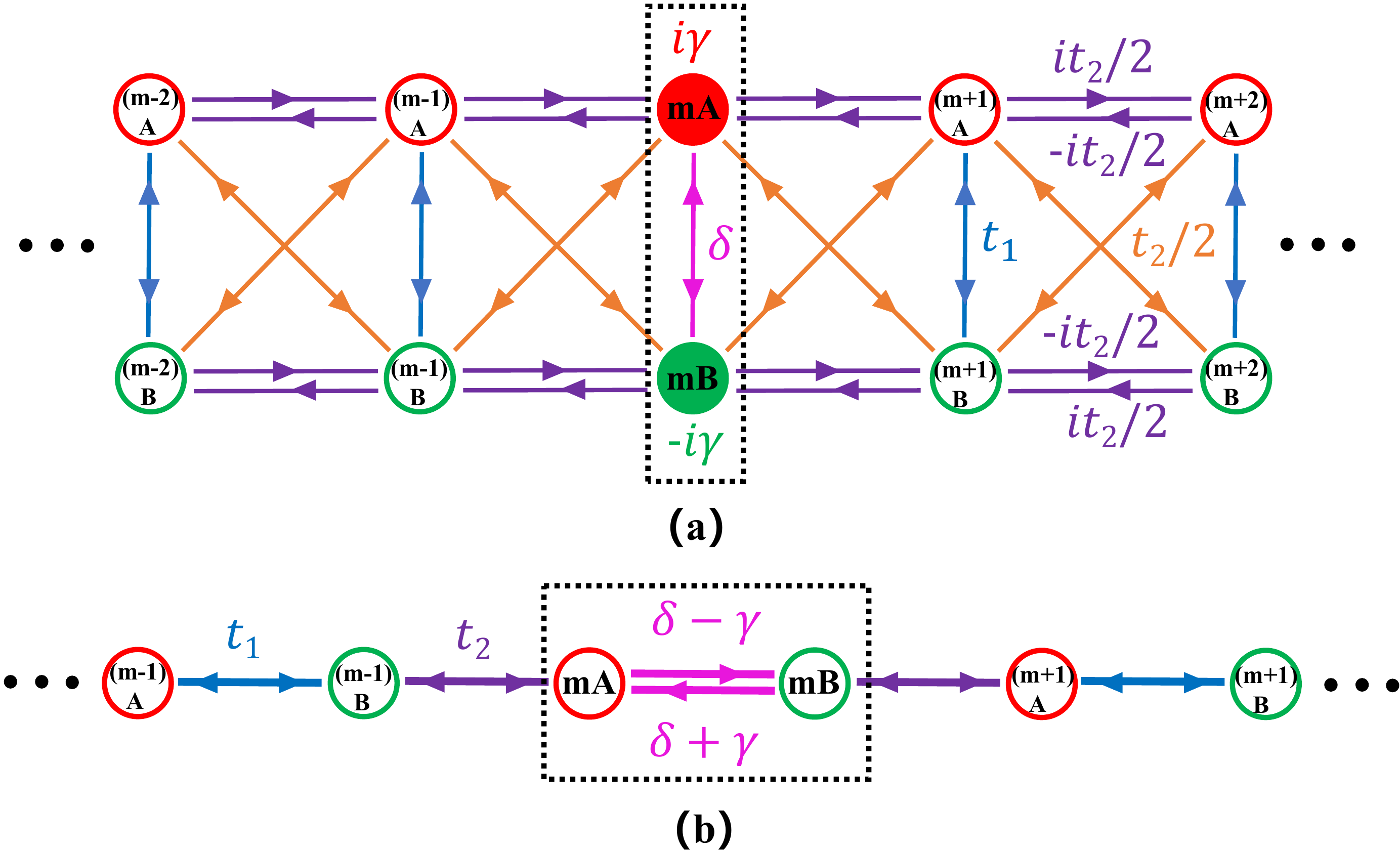}
\caption{(a) Sketch of the double chain model described in Eq. (\ref{H}). The red and green circles represent the $A$ and $B$ sublattice, respectively. The gain and loss terms are added only in the $m$th unit cell. (b) SSH model with nonreciprocal hopping on a single bond. The two models are related by a similarity transformation.}%
\label{fig1}
\end{figure}

\section{Scale-free localization induced by local non-Hermiticity}
\subsection{Models and solutions}
We start from a closed double chain model with an on-site gain/loss term of strength $\gamma\in\mathbb{R}$ residing on a single rung, as depicted in Fig. \ref{fig1}(a). The Hamiltonian is expressed as
\begin{equation}\label{H}
\begin{split}
\hat{H}_{dc}=&\hat{H}_0 + i\gamma \hat{c}_{mA}^{\dag
}\hat{c}_{mA}-i\gamma\hat{c}_{mB}^{\dag
}\hat{c}_{mB},
\end{split}
\end{equation}
where $m$ is the position of the impurity rung. $\hat{H}_0$ is the tight-binding Hermitian Hamiltonian described by
\begin{equation}
\begin{split}
\hat{H}_0=&t_{1}\sum\limits_{n\neq m}\left( \hat{c}_{nA}^{\dag
}\hat{c}_{nB}+h.c.\right) + \delta\left( \hat{c}_{mA}^{\dag
}\hat{c}_{mB}+h.c.\right)\\
&+\frac{t_{2}}{2}\sum\limits_{n=1}^{N}\left( \hat{c}_{n+1,B}^{\dag
}\hat{c}_{nA}+\hat{c}_{n+1,A}^{\dag
}\hat{c}_{nB}+h.c.\right)\\
&+\frac{t_{2}}{2}\sum\limits_{n=1}^{N}\left(i\hat{c}_{n+1,A}^{\dag
}\hat{c}_{nA}-i\hat{c}_{n+1,B}^{\dag
}\hat{c}_{nB}+h.c.\right).\\
\end{split}
\end{equation}
Here $\hat{c}_{n,A/B}^{\dag}~(\hat{c}_{n,A/B})$ is the particle creation (annihilation) operator at the $A/B$-sublattice of the $n$th cell, and $N$ is the number of unit cells. $t_1\in\mathbb{R}$ is the intracell hopping strength except for the $m$th rung with $\delta\in\mathbb{R}$. $t_2\in\mathbb{R}$ is the intercell hopping. The non-Hermiticity is introduced solely through the local gain/loss on the $m$th rung. For convenience, we set $\gamma>0, t_2>0, \delta>0$ without loss of generality and $t_1=1$ as energy unit.

The double chain model possesses $\mathcal{PT}$-symmetry \cite{PTS} $(\mathcal{PT})\hat{H}_{dc}(\mathcal{PT})^{-1}=\hat{H}_{dc}$ with $\mathcal{P}=\bigoplus_{n=1}^{N}\sigma_n^{x}$ and $\mathcal{T}$ the complex conjugate. $\hat{H}_{dc}$ also has a sublattice symmetry $\Gamma\hat{H}_{dc}\Gamma^{-1}=-\hat{H}_{dc}$ with $\Gamma=\bigoplus_{n=1}^{N}\sigma_n^{y}$. Here, $\sigma_n^{x}$ and $\sigma_n^{y}$ are Pauli matrices for the $n$th unit cell, $\bigoplus$ is the direct sum. A similarity transformation $S=\bigoplus_{n=1}^{N}S_{\sigma}^{n}$ with $S_{\sigma}^{n}=e^{i\frac{\pi}{4}\sigma_x^n}$ brings the Hamiltonian to the more familiar SSH model as depicted in Fig. \ref{fig1}(b).
The explicit form of the non-Hermitian SSH model is written as
\begin{equation}
\begin{split}
\hat{H}_{SSH}=&t_{1}\sum\limits_{n\neq m}\left( \hat{c}_{nB}^{\dag
}\hat{c}_{nA}+h.c.\right)+\\
& t_{2}\sum\limits_{n=1}^{N}\left(\hat{c}_{n+1,A}^{\dag
}\hat{c}_{nB}+h.c.\right) +
\\
& (\delta+\gamma) \hat{c}_{mA}^{\dag}\hat{c}_{mB}+ (\delta-\gamma)\hat{c}_{mB}^{\dag
}\hat{c}_{mA}.
\end{split}
\label{NHSSE}
\end{equation}
Clearly, the local gain/loss term is transformed to the nonreciprocal hopping term inside the $m$th unit cell. The symmetries ensure that the eigenvalues of $\hat{H}_{dc}$ and $\hat{H}_{SSH}$ appear in quartet of $(E,-E, E^{\ast},-E^{\ast})$ (See Appendix \ref{Appendix A}).
\begin{figure*}[tbp]
\includegraphics[width=0.98\textwidth]{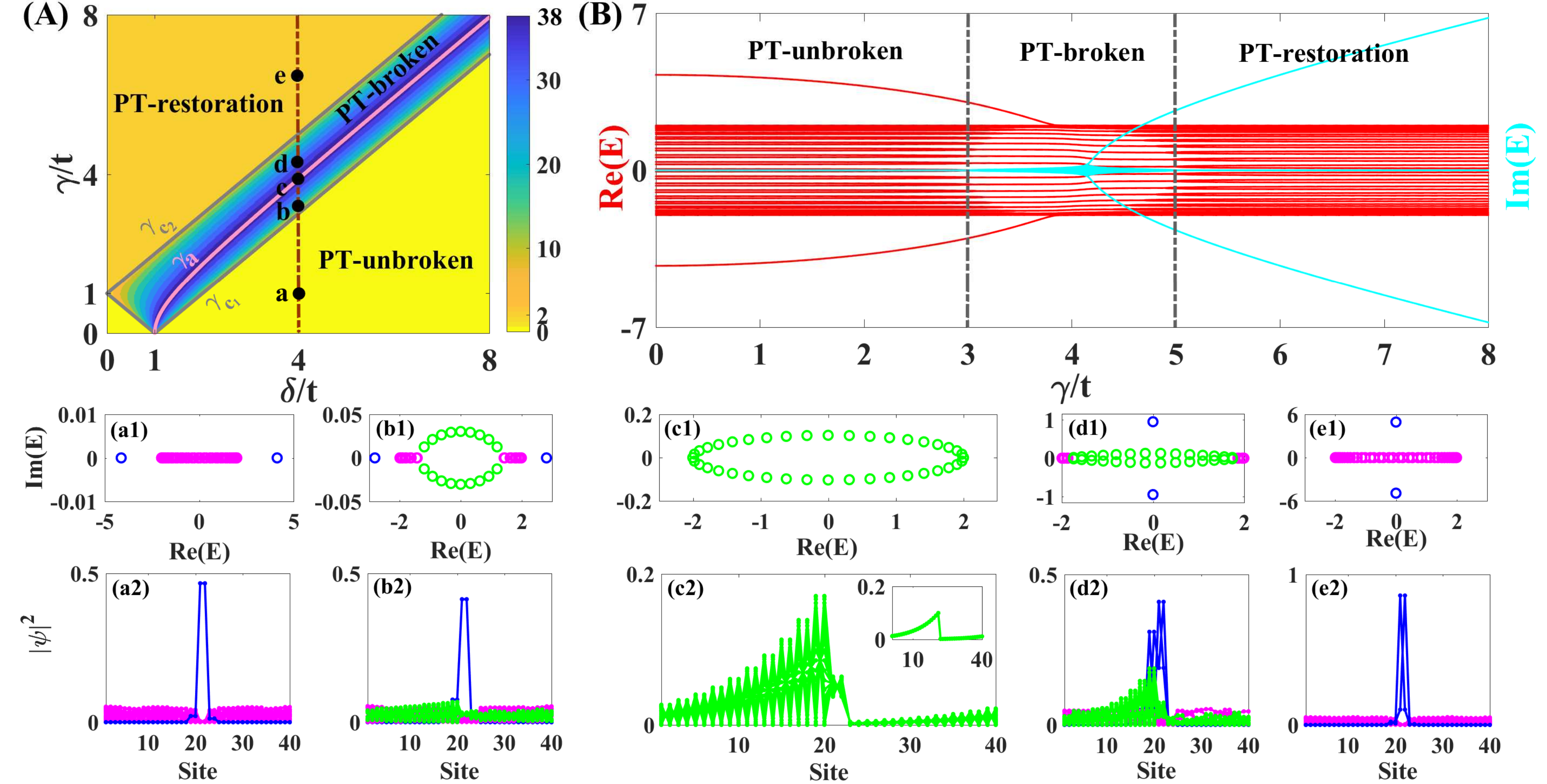}
\caption{(A) Phase diagram of the double chain model Eq. (\ref{H}). The phase boundaries (in gray lines) between the three regimes are given by Eq. (\ref{pb}). The number of complex eigenenergies is coded in colors for lattice size $2N=40$. At $\gamma=\gamma_a=\sqrt{\delta^2-t^2}$ (pink line), there are $N_{\mathrm{Im}}=2(N-1)$ complex eigenenergies and all eigenstates are scale-free localized (SFL). (B) Energy spectra with respect to $\gamma$ for $\delta=4t$ (the brown line in (A)). The real/imaginary parts of eigenenergies are marked in red/cyan. (a1-e1) Spectra on the complex-energy plane for $\gamma=1,~3.2,~\sqrt{15},~4.3,~6.5$, corresponding to dots 'a-e' in (A), respectively. (a2-e2) Spatial profiles of all eigenstates for the same parameters as (a1-e1). The inset in (c2) plots the wave function for the related SSH model $\hat{H}_{SSH}$. In (a1-e1,~a2-e2), the extended/bound/SFL states are marked in magenta/blue/green, respectively. The impurity rung is set at $m=11$.}
\label{fig2}
\end{figure*}

In the following, we focus on the double chain model $\hat{H}_{dc}$ and analyze its spectral properties. Using the method developed in Ref. \cite{GuoCX} (See Appendix \ref{Appendix B}), the eigenenergies can be expressed as
\begin{equation}\label{XSspectrum1}
E=\pm\sqrt{2t_{1}t_{2}\cos \theta+t_{1}^2+t_{2}^2},
\end{equation}
with the complex variable $\theta$ determined by the condition
\begin{equation}\label{eq-theta}
\sin[(N+1)\theta]+\eta_3\sin(N\theta)-\eta_2\sin[(N-1)\theta]-\eta_1\sin\theta=0.
\end{equation}
Here $\eta_1= \frac{2\delta}{t_{1}}$, $\eta_2=\frac{\delta^2-\gamma^2}{t_{1}^{2}}$, $\eta_3=\frac{t_{1}^2-\delta^2+\gamma^2}{t_{1}t_{2}}$. The eigenfunctions are
$\Psi=S^{-1}(...,\overline{\psi}_{n,A},\overline{\psi}_{n,B},...)$, taking the superposition form:
\begin{equation}
\begin{split}
\overline{\psi}_{n,A}&=\overline{c}_1e^{i(N-\tilde{n})\theta}\overline{\phi}_A^{(1)}+ \overline{c}_2e^{-i(N-\tilde{n})\theta}\overline{\phi}_A^{(2)},\\
\overline{\psi}_{n,B}&=\overline{c}_1e^{i(N-\tilde{n}+1)\theta}\overline{\phi}_B^{(1)}+ \overline{c}_2e^{-i(N-\tilde{n}+1)\theta}\overline{\phi}_B^{(2)}.
\end{split}
\end{equation}
Here $\tilde{n}$ is the distance from the impurity from the left side \cite{nn}. It is clear that the imaginary part of $\theta$ determines the localization properties of eigenfunctions. To grasp the main physics, we first consider the $t_2=t_1=t$ case where the eigenvalues reduce to
\begin{equation}
E=\pm2t\cos\frac{\theta}{2},
\end{equation}
and leave the discussions on generic cases to the Appendix \ref{Appendix C}.

\subsection{Sequential breaking of $\mathcal{PT}$-symmetry and spectral coalescence}
We investigate the evolution of energy spectra and the $\mathcal{PT}$-transition with respect to varying gain/loss strength $\gamma$ by solving Eq. (\ref{eq-theta}). The phase diagram is summarized in Fig. \ref{fig2}(A). There exist three distinct regimes, dubbed $\mathcal{PT}$-unbroken, $\mathcal{PT}$-broken, and $\mathcal{PT}$-restoration, respectively. Their boundaries are determined by
\begin{equation}\label{pb}
\gamma_{c_1}=|\delta-t|,~~~~\gamma_{c_2}=\delta+t,
\end{equation}
as marked in gray lines in Fig. \ref{fig2}(A). As an example, Fig. \ref{fig2}(B) plots the spectra versus $\gamma$ with fixed $\delta=4t$. When $\gamma<\gamma_{c_1}$, Eq. (\ref{eq-theta}) has $(N-1)$ real roots corresponding to $2(N-1)$ extended bulk states, and a purely imaginary root corresponding to a pair of real-energy bound states residing at the impurity rung [see Figs. \ref{fig2}(a1),(a2)]. In this regime, all eigenvalues are real, and the system is in the $\mathcal{PT}$-unbroken phase. Increasing $\gamma$ to exceed $\gamma_{c_1}$, $\theta$ starts to take complex roots and the corresponding eigenvalues become complex. The system enters into the $\mathcal{PT}$-broken phase. The number of real roots of $\theta$ shrinks first, and reaches its minimum at $\gamma=\gamma_a$ with $\gamma_a=\sqrt{\delta^2-t^2}$ [See the pink line in Fig. \ref{fig2}(A)] and then increases. The $\mathcal{PT}$-symmetry breakings start from the band center at $\mathrm{Re}(E)=0$ to the band edges for $\gamma_{c_1}<\gamma<\gamma_a$, through a sequence of exceptional points where two nearby real eigenvalues coalesce. For $\gamma_a<\gamma<\gamma_{c_2}$, two complex eigenvalues coalesce again and their eigenstates restore the $\mathcal{PT}$-symmetry. When $\gamma>\gamma_{c_2}$, we regain $(N-1)$ real roots and a complex root for $\theta$. They correspond to $2(N-1)$ extended bulk states of real eigenvalues and two bound states of purely imaginary eigenvalues, as shown in Figs. \ref{fig2}(e1),(e2). We dub this regime the $\mathcal{PT}$-restoration phase. In the $\mathcal{PT}$-unbroken regime with $\delta<t$, no bound states exist as Eq. (\ref{eq-theta}) has $N$ real roots. Notably at $\delta=t$, an arbitrarily small gain/loss or non-reciprocity induces the $\mathcal{PT}$-symmetry breakings and drastically changes all the eigenstates as will be discussed later.

\subsection{Scale-free localization}
\begin{figure}[tbp]
\includegraphics[width=0.46\textwidth]{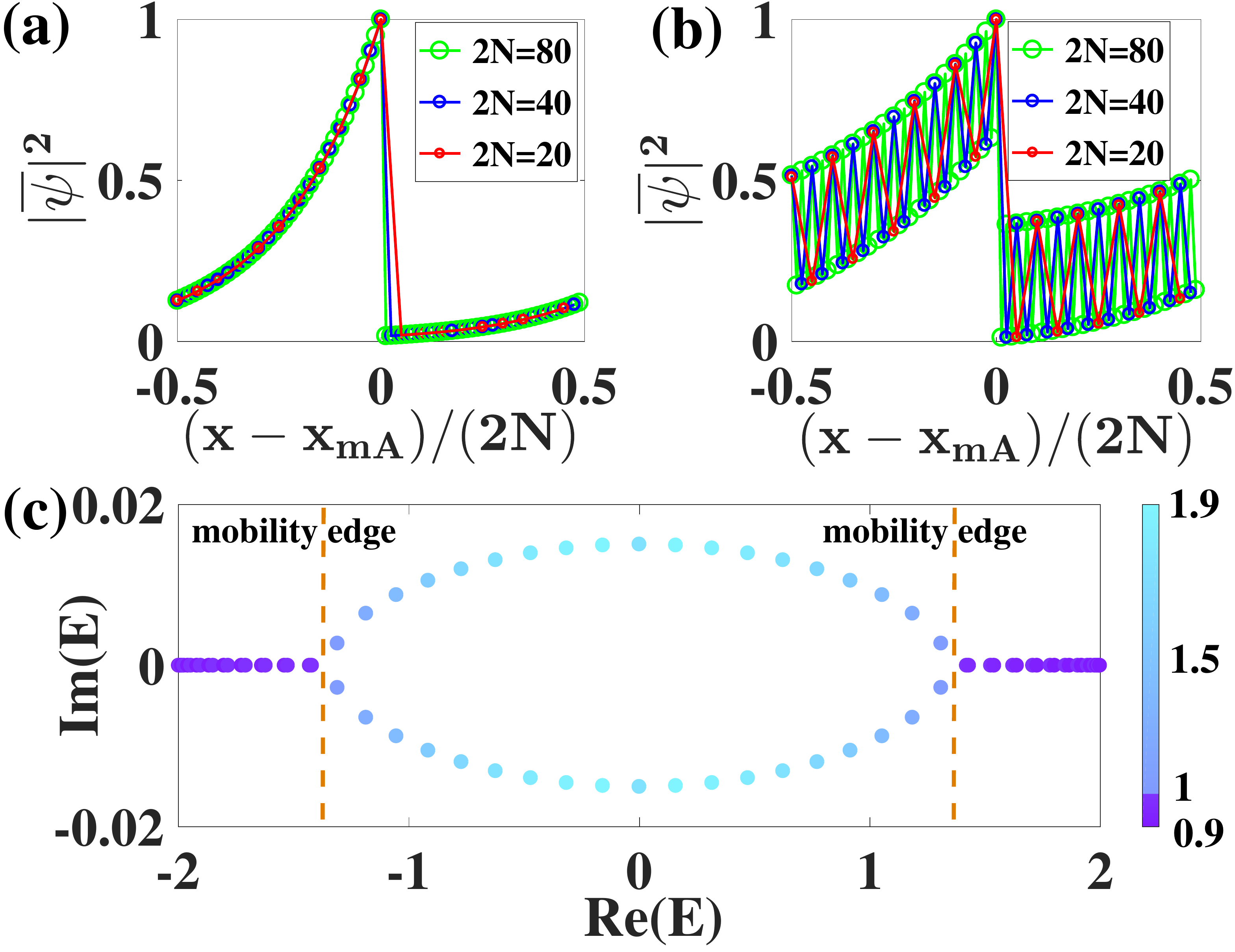}
\caption{(a) Rescaled spatial distributions of all eigenstates for the SSH model in Fig. \ref{fig1}(b) with $\gamma=\gamma_a$. (b) Rescaled spatial distribution of the eigenstate with the largest imaginary part of eigenvalue for various system sizes, $\gamma=3.4$. The localization lengths divided by system sizes are equal. (c) Mobility edges (dashed lines) extracted from the quantity $\chi$ for different eigenstates, $\gamma=3.2$, $2N=80$. For (a)-(c), $\delta=4t$, and the impurity resides at $m=N/2+1$ for various system sizes.}%
\label{fig3}
\end{figure}
We proceed to consider the spatial distributions of eigenstates in the $\mathcal{PT}$-broken regime $\gamma_{c_1}<\gamma<\gamma_{c_2}$, where complex eigenvalues (corresponding to complex roots of $\theta$) emerge. We start from the $\gamma=\gamma_a$ case. There are $2(N-1)$ complex eigenvalues and two real eigenvalues forming an oval on the complex-energy plane [See Fig. \ref{fig2}(c1)]. The $\theta$-solutions are $$\theta=\theta_R+i\theta_I=\frac{2l\pi}{N}-i\frac{\log\mu}{N}$$ with $\mu=\delta/t +\sqrt{(\delta/t)^2-1}$, $l=0,1,\cdots,N-1$. Therefore, we have $$|\mathrm{Im}(E)|=(\mu^{1/(2N)}-\mu^{-1/(2N)})|\sin\theta_R|\approx\frac{\log\mu}{N}|\sin\theta_R|.$$ Obviously, the local non-Hermitian term contributes a $1/N$-order correction to the imaginary part of the $\theta$-roots as well as the eigenenergies. Further, all eigenstates have the same spatial distributions. Formally, the moduli of all wave functions are
\begin{equation}
\begin{split}
|\overline{\psi}_{x}|&=
\left\{
  \begin{array}{ll}
    \mu^{\frac{x-x_{mA}}{2N}+1}, & \hbox{$x\leq x_{mA}$;}\\
    \mu^{\frac{x-x_{mA}}{2N}}, & \hbox{$x> x_{mA}$.}
  \end{array}
\right.
\end{split}
\end{equation}
Here $x_{mA}=2m-1$, $x=2n-1$ or $2n$ represent the A or B sublattice of the $n$th unit cell. The localization length of these wave functions is
\begin{equation}
\xi=\frac{2N}{\log \mu},
\end{equation}
which is proportional to the system size. As plotted in Fig. \ref{fig2}(c2), the spatial profiles of all eigenstates decay away from the impurity in a unidirectional way. The linear dependence of $\xi$ on the system size suggests that such unidirectional accumulation is the scale-free localization \cite{LiLH2021,Murakami}. Note the difference from the usual non-Hermitian skin modes of finite localization length independent of $N$. As a striking feature, the rescaled spatial profiles of SFL states (by the system size) stay intact varying system sizes, as depicted in Fig. \ref{fig3}(a).

The emergence of SFL states is not limited to the special parameter $\gamma=\gamma_a$. When $\gamma$ deviates from $\gamma_a$, the extended states associated with real eigenvalues and SFL states associated with complex eigenvalues coexist, as shown in Figs. \ref{fig2}(b1),(b2),(d1),(d2). Figure \ref{fig3}(b) further plots the rescaled profiles of a chosen complex-energy eigenstate for $\gamma\neq\gamma_a$. The scale-free localization accompanied by the $\mathcal{PT}$-symmetry breaking can be understood from the dispersion relation Eq. (\ref{XSspectrum1}). Heuristically, the local non-Hermitian term contributes a $1/N$-order correction to both the imaginary part of eigenenergies and roots of $\theta$ in the wave functions, yielding localization length of the order of system size $N$. In the $\mathcal{PT}$-broken regime, the extended and SFL states are separated by mobility edges. They can be distinguished by an \textit{ad hoc} quantity
\begin{equation}
\begin{split}
\chi=\frac{\sum_{x\in \mathrm{left}} |\overline{\psi}_{x}|^2}{\sum_{x\in \mathrm{right}} |\overline{\psi}_{x}|^2},
\end{split}
\end{equation}
where $x\in \mathrm{left/right}$ labels lattice sites on the left/right half side of the impurity. The positions of mobility edges can be read out from the discontinuity of $\chi$ (for extended states, $\chi\approx 1$ and for SFL states $\chi>1$), as shown in Fig. \ref{fig3}(c). For the general case of $t_1 \neq t_2$, SFL states appear after the $\mathcal{PT}$-transition. A full scale-free localization of all eigenstates occurs when $\gamma= \sqrt{\delta^2-t_1^2}$ (See Appendix \ref{Appendix C}).

\section{Generality of SFL states}
The above double-stranded or SSH model is for illustrative purposes. Roughly, the $\mathcal{PT}$-symmetry imposes a threshold of the strength of non-Hermiticity to induce scale-free localization. Yet, we emphasize that the local non-Hermiticity-induced scale-free localization is a general phenomenon. It exists in a much broader context, regardless of the $\mathcal{PT}$-symmetry and coalescence of extended eigenstates or even a priori Bloch-band description of the underlying Hermitian system.
In this section, we demonstrate that the SFL states can be induced by a single lossy impurity and may survive even when incommensurate lattice potential is added.

\subsection{Scale-free localization induced by local on-site imaginary potential}
We consider the model of a closed chain with the local non-Hermiticity given by an imaginary on-site potential (a single lossy impurity).
Explicitly, the Hamiltonian of the single-impurity model is given by
\begin{equation}
\hat{H} =\sum_{n=1}^{L} \left[ t (\hat{c}_{n+1}^{\dag
}\hat{c}_{n} + h.c.)+i\gamma \hat{c}_{m}^{\dag}\hat{c}_{m} \right].
\label{impurity}
\end{equation}
This model can be analytically solved by following the same method in Refs. \cite{GuoCX,YXLiu}. (The detailed derivation is given in Appendix \ref{Appendix D}.) The eigenvalues are given by
\begin{equation}
E=2t \cos \theta,
\end{equation}
where $\theta$ is determined by
\begin{equation}\label{Seq-thetaAA}
\sin(L\theta/2)\left[2t \sin \theta \sin(L \theta/2)+ i \gamma \cos(L \theta/2) \right]=0.
\end{equation}
The wave function of the single-impurity model can be obtained as $\Psi=(\psi_{1},\psi_{2},\cdots,\psi_{m},\cdots,\psi_{L-1},\psi_{L})^T$  with the superposition form:
\begin{equation}
\begin{split}
\psi_{n}=
\left\{
  \begin{array}{ll}
    c_{1}e^{i(L-m+n)\theta}+c_{2}e^{-i(L-m+n)\theta},  \hbox{$1\leq n\leq j$;}\\
    c_{1}e^{i(n-m)\theta}+c_{2}e^{-i(n-m)\theta}, ~~~ \hbox{$j< n\leq L$.}
  \end{array}
\right.
\end{split}
\end{equation}
Obviously, the localization properties of eigenfunctions are determined by the imaginary part of $\theta$.
There are two types of solutions for the Eq. (\ref{Seq-thetaAA}). The first type is from
\begin{equation}\label{Sspectrum1AA1}
\sin(L\theta/2)=0.
\end{equation}
The roots of Eq. (\ref{Sspectrum1AA1}) are $\theta=\frac{2l\pi}{L}$ with $l=1,2,\cdots,L/2-1$ for even $L$, and $l=1,2,\cdots,(L-1)/2$ for odd $L$. Thus there are $L/2-1$ real eigenenergies for $L\in \mathrm{even}$, and $(L-1)/2$ real eigenenergies for $L\in \mathrm{odd}$. Their corresponding eigenstates with odd-parity are all extended and unaffected by the local on-site imaginary potential.

\begin{figure}[htb]
\includegraphics[width=0.48\textwidth]{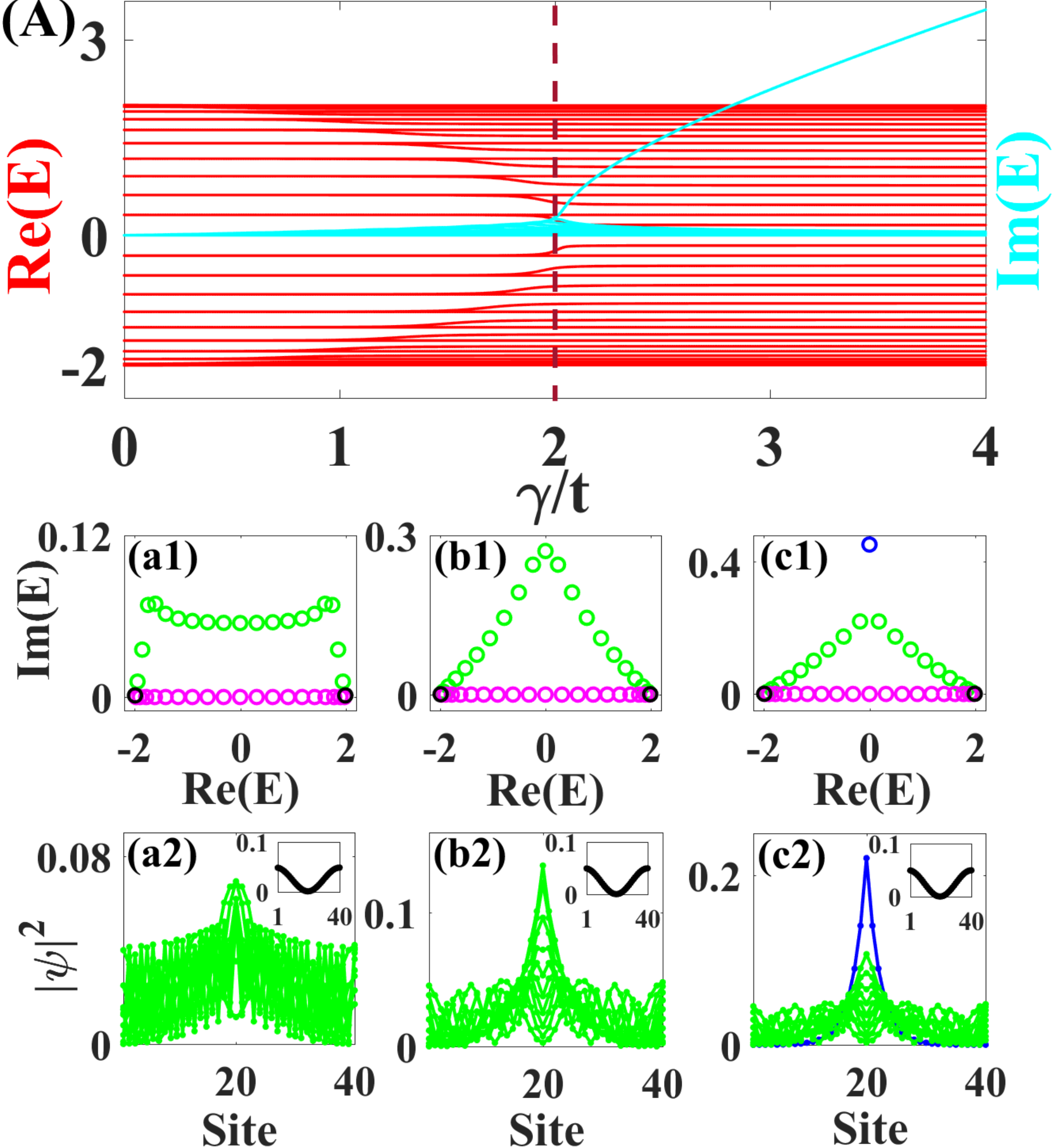}
\caption{(A) Energy spectra versus $\gamma/t$ for the single-impurity model on a closed chain. Red/cyan curves lines represent real/imaginary parts of eigenenergies. The dotted brown line separates the two regions with or without a bound state. (a1)-(c1) Energy spectra on the complex energy plane for the single-impurity model with $\gamma=1,~2,~2.05$, respectively. (a2)-(c2) Spatial distributions of the chosen eigenstates marked by green circles in (a1)-(c1). The insets in (a2)-(c2) plot the wave functions of the chosen eigenstates marked by black circles in (a1)-(c1). The parameters are chosen as $L=40$, $m=20$, $t=1$.}%
\label{fig6}
\end{figure}

The other eigenstates come from the second type of solutions:
\begin{equation}\label{Sspectrum1AA2}
2t \sin \theta \sin(L \theta/2)+ i \gamma \cos(L \theta/2)=0,
\end{equation}
and the corresponding roots $\theta$ are complex denoted as $\theta=\theta_R+i\theta_I$. Equation (\ref{Sspectrum1AA2}) has a root with $\theta_I\propto L^0$ only if $\gamma>2t$. Explicitly, the solution is written as
\begin{equation}
\theta=\frac{\pi}{2}+i\mathrm{arcosh}\big(\frac{\gamma}{2t}\big),
\end{equation}
which is associated with a bound state. In short, there is only a bound state with $\theta=\frac{\pi}{2}+\mathrm{arcosh}\big(\frac{\gamma}{2t}\big)$ when $\gamma>2t$, as shown in Fig. \ref{fig6}.
For the other complex roots of Eq. (\ref{Sspectrum1AA2}), their imaginary part satisfy $\theta_I\propto\frac{1}{L}$. In another word, the localization length of these eigenstates (except for bound states) with complex roots is proportional to the system size $\xi\propto L$, and these eigenstates are SFL states.

In Fig. \ref{fig6}(A), we show the energy spectra of the single-impurity model as $\gamma/t$ varies. A bound state appears when $\gamma>2t$ as expected. The energy spectra and spatial profiles of the eigenstates with $\gamma=1,~2,~2.05$ are plotted respectively in Figs. \ref{fig6}(a1)-(c1) and \ref{fig6}(a2)-(c2). Clearly, nearly half of all eigenstates are SFL or extended, consistent with our exact solutions. Almost all of the SFL states accumulate around the impurity, except for a pair of states with a very small imaginary part of eigenvalue, as displayed in the insets of Figs. \ref{fig6}(a2)-(c2). Therefore, for the single-impurity model, only half of all eigenstates determined by Eq. (\ref{Sspectrum1AA2}) are affected by the impurity, and the other half determined by Eq. (\ref{Sspectrum1AA1}) are irrelevant to the impurity strength due to their odd parity. Except for a bound state, the other even-parity states become SFL states.
\begin{figure}[tbh]
\includegraphics[width=0.46\textwidth]{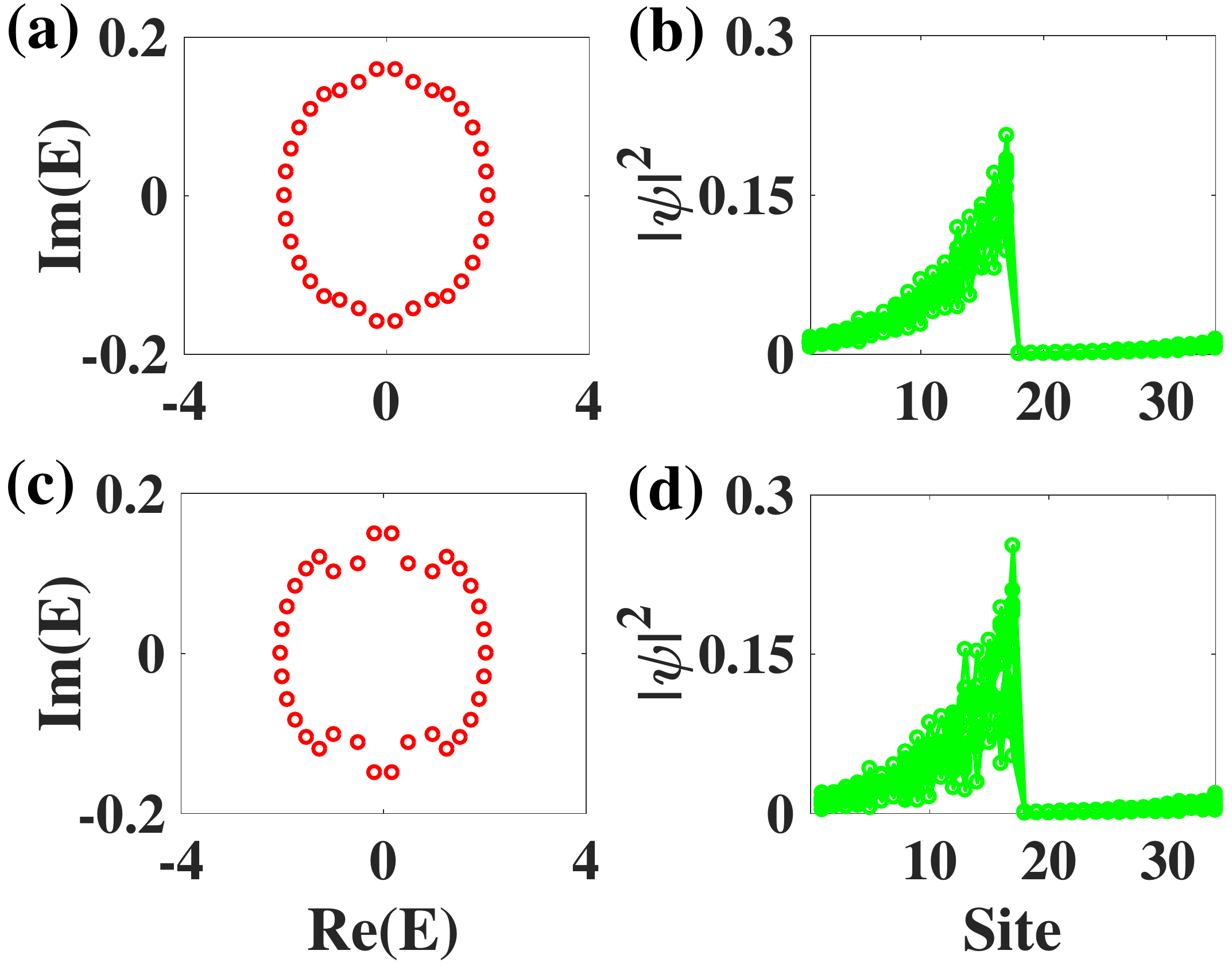}
\caption{(a)(b) Energy spectra and spatial distributions of all the eigenstates for the non-Hermitian AA model with local nonreciprocal hopping for $\lambda=0.1$. (c)(d) Energy spectra and spatial distributions of all the eigenstates for the non-Hermitian AA model with local nonreciprocal hopping for $\lambda=0.2$. Common parameters: $L=34$, $m=17$, $t=1$, $\delta=8$, $\gamma=\sqrt{\delta^2-t^2}=\sqrt{63}$.}
\label{fig4}
\end{figure}
\begin{figure}[b]
\includegraphics[width=0.46\textwidth]{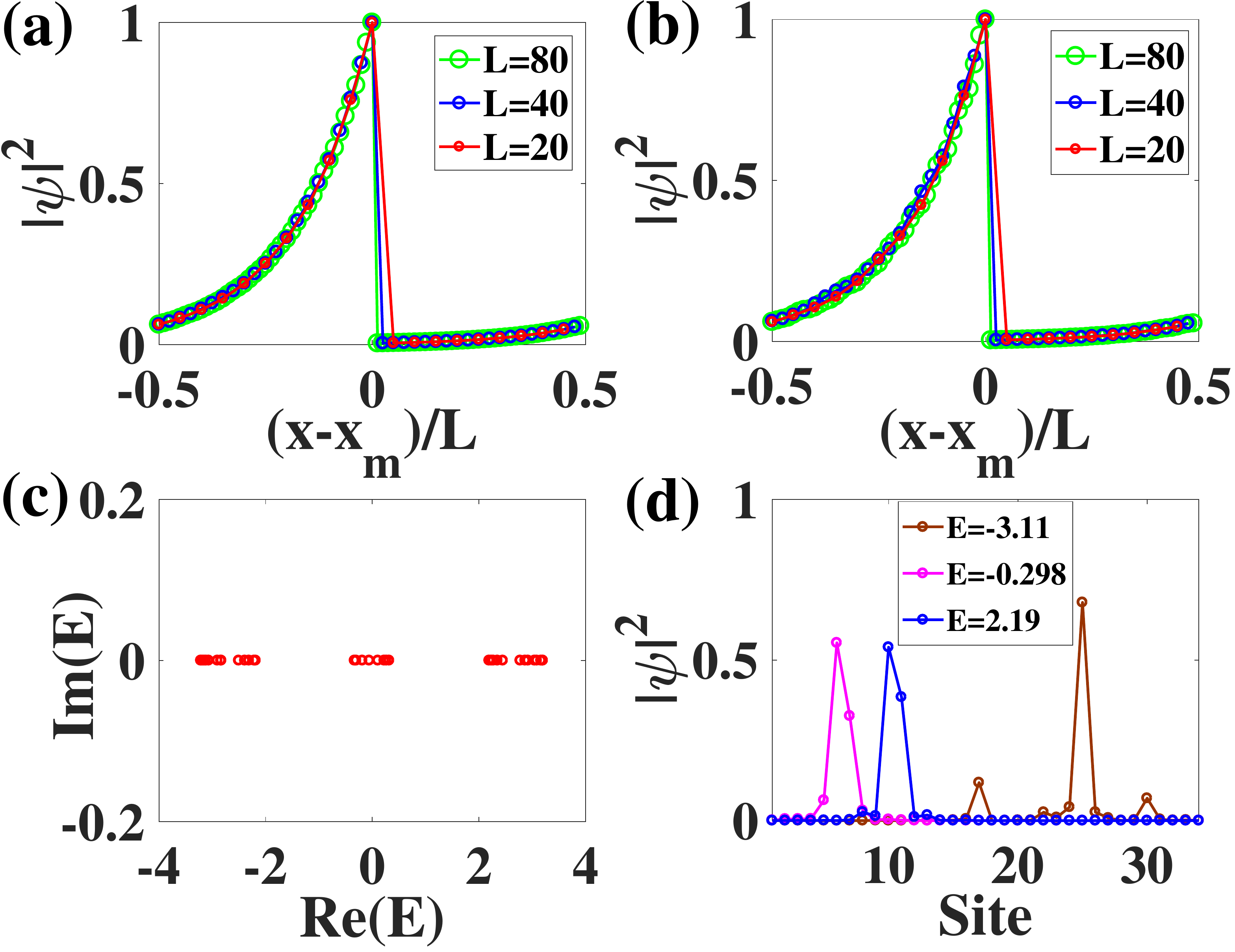}
\caption{(a)(b) Rescaled spatial distributions of the eigenstates with the largest imaginary part of eigenenergies for the non-Hermitian AA model with local nonreciprocal hopping. The system size takes $L=20,40,80$, and the impurity is located at $m=L/2+1$. (a) $\lambda=0.05$; (b) $\lambda=0.1$. (c)(d) Energy spectra and spatial distributions of several chosen eigenstates for the non-Hermitian AA model with local nonreciprocal hopping for $\lambda=1.4$, $L=34$, $m=17$. Common parameters: $t=1$, $\delta=8$, $\gamma=\sqrt{\delta^2-t^2}=\sqrt{63}$.}%
\label{fig5}
\end{figure}

\subsection{The non-Hermitian AA model with local nonreciprocal hopping}
Now we demonstrate that the SFL states can survive even when incommensurate lattice potential is added.
To be explicit, we consider the 1D quasiperiodic lattice described by the non-Hermitian AA model with a local non-Hermitian term. The Hamiltonian is
\begin{equation}\label{AAmodel1}
\hat{H}=\hat{H}_{AA}+\hat{H}_{NH},
\end{equation}
with
\begin{equation}
\hat{H}_{AA}=\sum_{n=1}^{L} \left[ t (\hat{c}_{n+1}^{\dag
}\hat{c}_{n} + h.c.)+2 \lambda \cos(2\pi\alpha n) \hat{c}_{n}^{\dag
}\hat{c}_{n} \right],
\end{equation}
here $\alpha=(\sqrt{5}-1)/2$ is an irrational number. In the following, we discuss the case with different forms of the local non-Hermitian term separately by using examples such as nonreciprocal hopping or imaginary on-site potential.

Here, we consider the non-Hermitian AA model of (\ref{AAmodel1}) with the local non-Hermiticity given by nonreciprocal hopping
\begin{equation}
\begin{split}
\hat{H}_{NH}&=(\delta+\gamma-t) \hat{c}_{m}^{\dag}\hat{c}_{m+1}+ (\delta-\gamma-t)\hat{c}_{m+1}^{\dag}\hat{c}_{m}.
\end{split}
\end{equation}
The case of $\lambda=0$ has been studied in Fig. \ref{fig2}. Note that the AA model $\hat{H}_{AA}$ undergoes a delocalization/localization phase transition at the critical strength of quasiperiodic potential $\lambda=t$ \cite{Aubry1980}.

We expect the SFL states to survive in the delocalization regime, for which all the unperturbed eigenstates of $\hat{H}_{AA}$ are extended. In particular, when $\gamma=\sqrt{\delta^2-t^2}$, a full scale-free localization persists for all eigenstates even in the presence of on-site incommensurate potential, as verified by our numerical results in Fig. \ref{fig4}. The eigenenergies and their associated SFL eigenstates which decay away from the impurity are shown in Figs. \ref{fig4}(a),(b) with $\lambda=0.1$, and in Figs. \ref{fig4}(c),(d) with $\lambda=0.2$. In Figs. \ref{fig5}(a),(b), we further plot the rescaled spatial profiles of the eigenstate with the largest imaginary part of eigenenergies for various system sizes. They coincide with each other, indicating their scale-free nature.

In contrast, when $\lambda$ lies in the localized regime, the eigenstates of the non-Hermitian AA model are insensitive to the non-Hermiticity, and no SFL state is observed. As shown in Figs. \ref{fig5}(c),(d) with a large incommensurate potential $\lambda=1.4$, all eigenstates become localized.

\begin{figure}[tbh]
\includegraphics[width=0.45\textwidth]{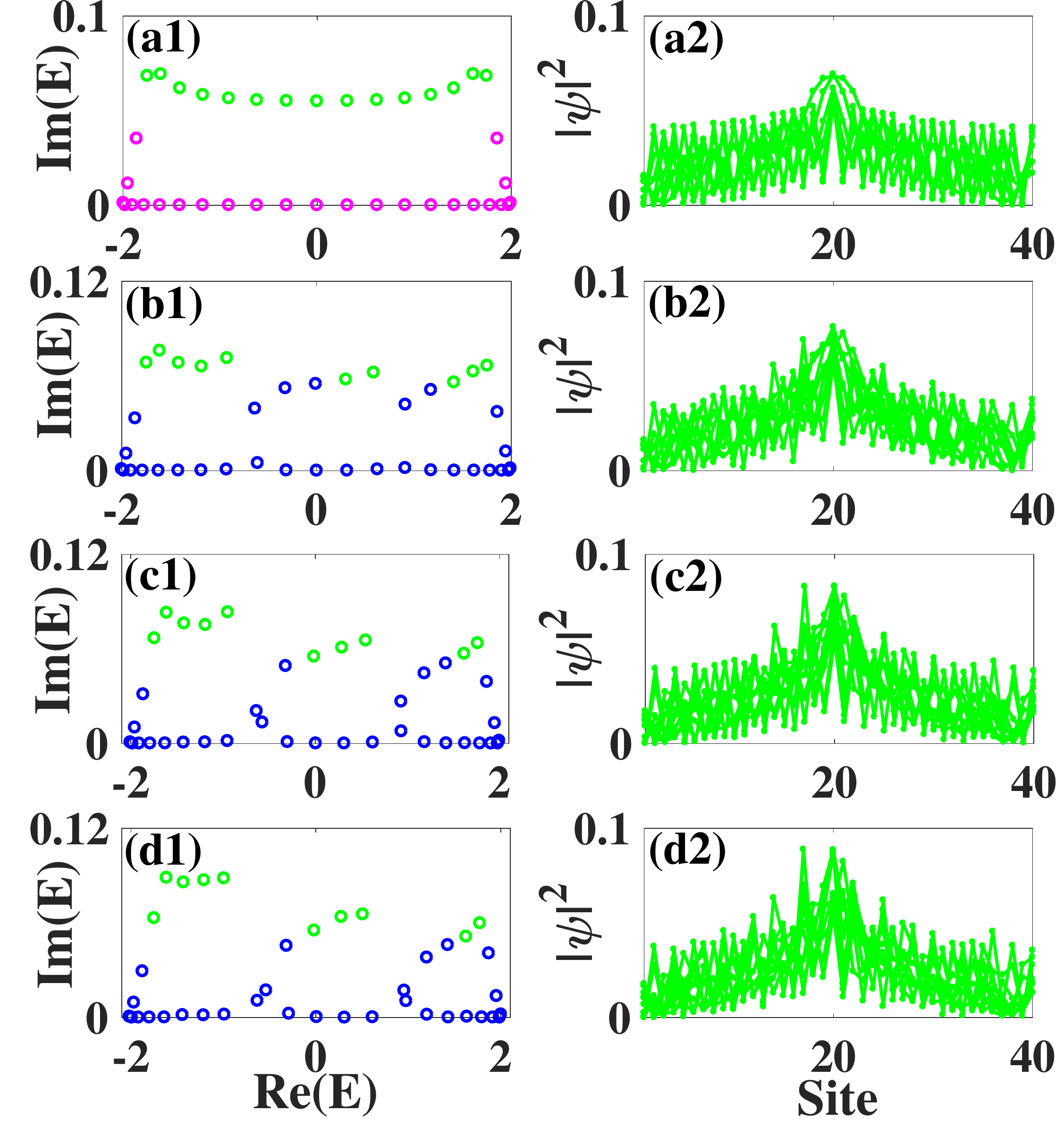}
\caption{(a1)-(d1) Energy spectra of the non-Hermitian AA model with local on-site imaginary potential for $\lambda=0,~0.05,~0.1,~0.15$, respectively. (a2)-(d2) Spatial distributions of the ten chosen eigenstates with largest imaginary parts of eigenenergies (marked by green dots in (a1)-(d1)) for $\lambda=0,~0.05,~0.1,~0.15$, respectively. Common parameters: $L=40$, $m=20$, $t=1$, $\gamma=1$.}%
\label{fig7}
\end{figure}
\begin{figure}[tbh]
\includegraphics[width=0.48\textwidth]{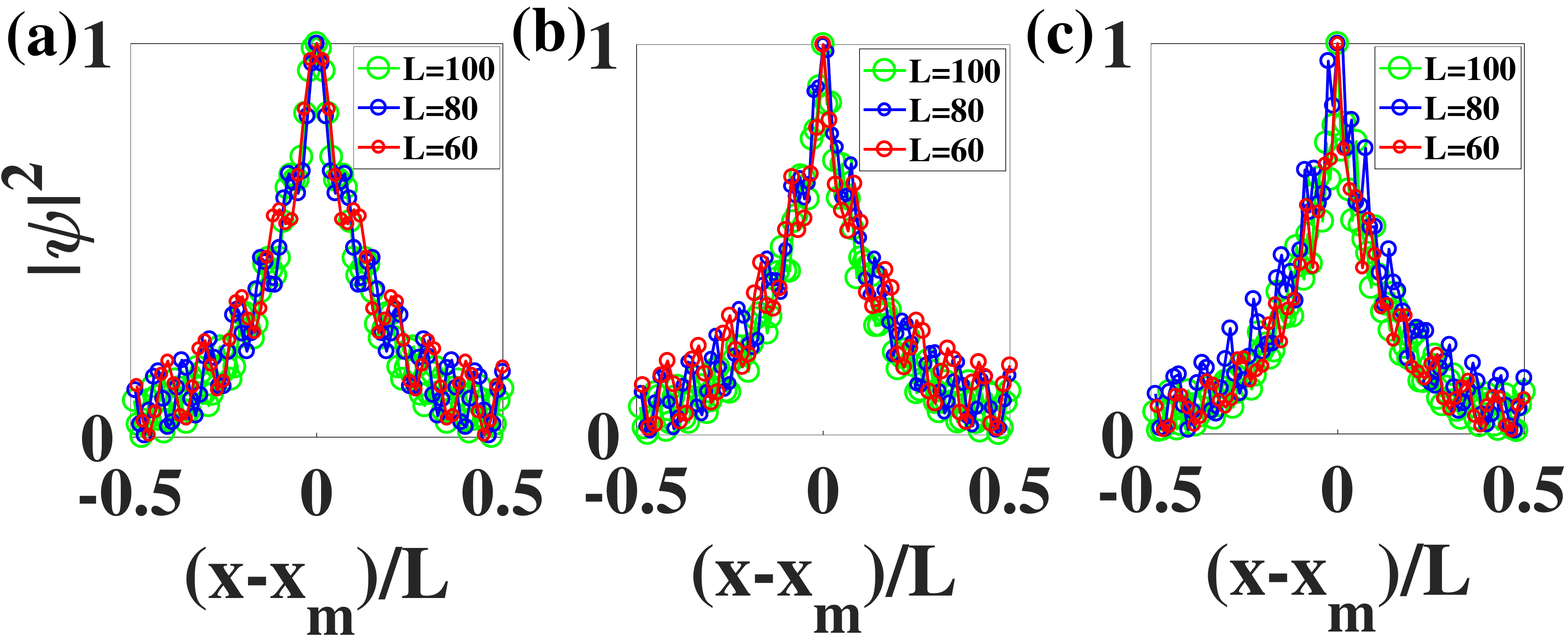}
\caption{Rescaled spatial distributions of the eigenstate with the largest imaginary part of eigenenergy for the non-Hermitian AA model with an on-site imaginary potential. The system sizes take $L=60,80,100$. (a) $\lambda=0$; (b) $\lambda=0.05$; (c) $\lambda=0.1$. Other parameters are $m=L/2$, $t=1$, $\gamma=1$.}%
\label{fig8}
\end{figure}

\subsection{The non-Hermitian AA model with local on-site imaginary potential}

Next we consider the non-Hermitian AA model of (\ref{AAmodel1}) with the local non-Hermiticity described by the local on-site imaginary potential, i.e.,
\begin{equation}
\hat{H}_{NH}=i\gamma \hat{c}_{m}^{\dag}\hat{c}_{m}.
\end{equation}
The case of $\lambda=0$ reduces to the model of (\ref{impurity}). For $\lambda \neq 0$, we can numerically diagonalize the Hamiltonian.
Our numerical results verify that the SFL states can survive for small incommensurate potential strength $\lambda$. In Fig. \ref{fig7}, we present the numerical results for four typical $\lambda$ (i.e., the strength of incommensurate potential) for the non-Hermitian AA model with imaginary on-site potential. We can clearly see that the SFL states survive a finite $\lambda$. Then we show the rescaled spatial distributions in Fig. \ref{fig8} for the specific eigenstate with the largest imaginary of eigenvalue. The spatial distributions for different system sizes almost coincide with each other, indicating that they are SFL states.

\section{Conclusions and Discussions}
To summarize, we have unveiled the emergence of local non-Hermiticity-induced SFL states by presenting exact solutions for the double chain model. The non-Hermitian term drives the system through a sequence of $\mathcal{PT}$-symmetry breakings, accompanied by the appearance of complex eigenenergies and SFL states. Mobility edges separate the residual extended states and SFL states till a full scale-free localization occurs. We have further demonstrated the generality and robustness of local non-Hermiticity-induced scale-free localization regardless of, e.g., the $\mathcal{PT}$-symmetry and the incommensurate lattice potential.

The lattice Hamiltonian with local non-Hermiticity (including both the non-reciprocity and gain/loss) should be readily implemented in various classical/quantum simulation platforms like electric circuits \cite{ExpCircuitImhof,ExpCircuitLiu}, optical \cite{ExpOptLonghi,ExpOptWeid,ExpOptZhu} or acoustic cavities \cite{ExpAcoZhang}, quantum walks \cite{ExpOptXue1,ExpOptXue2,ExpOptXue3}, and cold atoms \cite{ExpColdLapp,ExpColdLiang,ExpColdRen,ExpColdWYi}. The induced scale-free localization could thus be identified through the spectral measurement and spatial distributions of eigenstates in these platforms. Our results indicate that local non-Hermiticity drastically alters the bulk spectral properties. The next step is to investigate its influence on the macroscopic observables, phase transitions, and dynamical properties. Other important issues include extending the studies to higher dimensions and continuum systems (without lattices), and exploring the intriguing interplay between local non-Hermiticity, long-ranged couplings, many-body interactions, and other localization mechanisms.

\begin{acknowledgments}
We thank L. Li for helpful discussions. This work is supported by National Key Research and Development Program of China (Grant No. 2021YFA1402104 and Grant No. 2022YFA1405800), the NSFC under Grants No. 11974413, No. 12174436 and No. T2121001, and the Strategic Priority Research Program of Chinese Academy of Sciences under Grant No. XDB33000000. H. H. is also supported by the start-up grant of IOP, CAS.
\end{acknowledgments}

\appendix

\section{Symmetry analysis of the non-Hermitian SSH model}\label{Appendix A}
The non-Hermitian Su-Schrieffer-Heeger (SSH) model (\ref{NHSSE}) is related to the double chain model through the similarity transformation $S\hat{H}_{dc}S^{-1}=\hat{H}_{SSH}$. The $\mathcal{PT}$-symmetry and sublattice symmetry of the double chain model correspond respectively to pseudo-Hermitian symmetry and sublattice symmetry of the non-Hermitian SSH model. The pseudo-Hermitian symmetry which guarantees that the eigenvalues appear in $(E,E^{\ast})$ pair takes
\begin{equation}
\eta\hat{H}_{SSH}\eta^{-1}=\hat{H}_{SSH}^{\dag},
\end{equation}
where
\begin{equation}
\eta=\bar{I}_{L_1\times L_1} \bigoplus \bar{I}_{L_2\times L_2},
\end{equation}
and $\bar{I}_{L_{1/2}\times L_{1/2}}$ denotes the $L_{1/2}\times L_{1/2}$ matrix whose subdiagonal entries are all 1 and other entries are 0. Explicitly, we have $L_1=2(2m-1)$, $L_2=2(N-2m+1)$ if $m\leq \frac{N+1}{2}$ while $L_1=2(2m-N-1)$, $L_2=2(2(N-m)+1)$ for $m > \frac{N+1}{2}$. The sublattice symmetry reads
\begin{equation}
\overline{\Gamma}\hat{H}_{SSH}\overline{\Gamma}^{-1}=-\hat{H}_{SSH}
\end{equation}
with $\overline{\Gamma}=\bigoplus_{n=1}^{N}\sigma_z^{n}$. It ensures the eigenvalues appear in $(E,-E)$ pair. Therefore, the eigenvalues of $\hat{H}_{SSH}$ and $\hat{H}_{dc}$ appear in quartet of $(E,-E, E^{\ast},-E^{\ast})$.

\section{Exact solutions of the double chain model} \label{Appendix B}
\subsection{Solutions for the generic case}
Here we detail the exact solutions of the two models $\hat{H}_{dc}$ and $\hat{H}_{SSH}$ depicted in Fig. \ref{fig1}. For the double chain model, the eigenvalue equation is
\begin{equation}
\hat{H}_{dc}|\Psi\rangle=E|\Psi\rangle,
\end{equation}
with
\begin{equation}
|\Psi\rangle=\sum\limits_{n=1}^{N} (\psi_{n,A} \hat{c}_{nA}^{\dag } + \psi_{n,B} \hat{c}_{nB}^{\dag } ) |0\rangle.
\end{equation}
In its components,
\begin{equation}
\Psi=(\psi_{1A},\psi_{1B},\psi_{2A},\cdots,\psi_{mA},\psi_{mB},\cdots,\psi_{NA},\psi_{NB})^T.
\end{equation}
For the non-Hermitian SSH model, the eigenvalue equation is
\begin{equation}
\hat{H}_{SSH}|\overline{\Psi}\rangle=\overline{E}|\overline{\Psi}\rangle,
\end{equation}
where
\begin{equation}
\overline{\Psi}=(\overline{\psi}_{1A},\overline{\psi}_{1B},\overline{\psi}_{2A},\cdots,\overline{\psi}_{mA},\overline{\psi}_{mB},\cdots,\overline{\psi}_{NA},\overline{\psi}_{NB})^T.
\end{equation}
As mentioned in the main text, the two models are related by a similarity transformation. Hence they have the same energy spectra $E=\overline{E}$, with their wave functions related by the similar transformation, i.e.,
\begin{equation}
|\Psi\rangle=S^{-1}|\overline{\Psi}\rangle.
\end{equation}

In the following, we focus on the non-Hermitian SSH model and obtain its solutions. Formally, for the bulk lattice sites, the eigenvalue equation takes
\begin{equation}
t_{1}\overline{\psi} _{sA}-E\overline{\psi} _{sB}+t_{2}\overline{\psi} _{s+1,A}=0,\label{XSbk1}\\
\end{equation}
with $s=1,\cdots,m-1,m+1,\cdots,N$, and
\begin{equation}
t_{2}\overline{\psi} _{sB}-E\overline{\psi} _{s+1,A}+t_{1}\overline{\psi} _{s+1,B}=0,\label{XSbk2}
\end{equation}
with $s=1,\cdots,m-2,m,\cdots,N$. For the impurity site, we have
\begin{eqnarray}
(\delta-\gamma)\overline{\psi} _{mA}-E\overline{\psi} _{mB}+t_{2}\overline{\psi} _{m+1,A} &=&0,\label{XSbd3}\\
t_{2}\overline{\psi} _{m-1,B}-E\overline{\psi} _{mA}+(\delta+\gamma)\overline{\psi} _{mB} &=&0.\label{XSbd4}
\end{eqnarray}
We take an ansatz wave function satisfying the bulk Eqs. (\ref{XSbk1})(\ref{XSbk2}) as follows:
\begin{equation}\label{XSFii}
\begin{split}
\overline{\Psi} _{i}=&(z_{i}^{N-m+1}\overline{\phi} _{A}^{(i)},z_{i}^{N-m+2}\overline{\phi} _{B}^{(i)},z_{i}^{N-m+2}\overline{\phi} _{A}^{(i)},z_{i}^{N-m+3}\overline{\phi} _{B}^{(i)},\\
&\cdots,z_{i}^{N}\overline{\phi} _{A}^{(i)},z_{i}\overline{\phi} _{B}^{(i)},\cdots,z_{i}^{N-m}\overline{\phi} _{A}^{(i)},z_{i}^{N-m+1}\overline{\phi} _{B}^{(i)})^T.
\end{split}
\end{equation}
Inserting the ansatz into Eqs. (\ref{XSbk1}),(\ref{XSbk2}) yields the expression of eigenvalue in terms of $z_i$:
\begin{equation}\label{XSEz1}
E = \pm \sqrt{\frac{t_{1}t_{2}}{z_{i}}%
+t_{1}t_{2}z_{i}+t_{1}^2+t_{2}^2},
\end{equation}
and the relation between $\overline{\phi} _{A}^{(i)}$ and $\overline{\phi} _{B}^{(i)}$:
\begin{eqnarray}\label{XSFAB}
\overline{\phi} _{B}^{(i)}=\frac{E}{(t_{2}+t_{1}z_{i})}\overline{\phi} _{A}^{(i)}=\frac{(t_{1}+t_{2}z_{i})}{Ez_{i}}\overline{\phi} _{A}^{(i)}.
\end{eqnarray}
Obviously, there are two solutions $z_i$ (denoted as $z_1, z_2$) for a given $E$ from Eq. (\ref{XSEz1}) satisfying the constraint:
\begin{eqnarray}\label{XSz1z2}
z_{1}z_{2} &=&1.
\end{eqnarray}

The eigenfunction in general takes the form of the superposition:
\begin{equation}
\begin{split}
\overline{\Psi}=&\overline{c}_{1}\overline{\Psi} _{1}+\overline{c}_{2}\overline{\Psi} _{2}\\
\equiv& (\overline{\psi} _{1A},\overline{\psi} _{1B},\overline{\psi} _{2A}, \cdots,\overline{\psi} _{mA},\overline{\psi} _{mB}, \cdots, \overline{\psi} _{NA},\overline{\psi} _{NB})^{T},~~~ \label{XSWave}
\end{split}
\end{equation}
where
\begin{equation}
\begin{split}
\overline{\psi}_{n,A}&=
\left\{
  \begin{array}{ll}
    \sum_{i=1}^{2}(\overline{c}_{i}z_{i}^{N-m+n}\overline{\phi} _{A}^{(i)}),  \hbox{$1\leq n\leq m$;} \label{XSwave1}\\
    \sum_{i=1}^{2}(\overline{c}_{i}z_{i}^{n-m}\overline{\phi} _{A}^{(i)}),  \hbox{$m< n\leq N$;}
  \end{array}
\right.
\\
\overline{\psi}_{n,B}&=
\left\{
  \begin{array}{ll}
    \sum_{i=1}^{2}(\overline{c}_{i}z_{i}^{N-m+1+n}\overline{\phi} _{B}^{(i)}),  \hbox{$1\leq n< m$;} \\
    \sum_{i=1}^{2}(\overline{c}_{i}z_{i}^{n-m+1}\overline{\phi} _{B}^{(i)}),  \hbox{$m\leq n\leq N$.}
  \end{array}
\right.
\end{split}
\end{equation}
Further substituting Eq. (\ref{XSWave}) into the impurity conditions Eqs. (\ref{XSbd3}),(\ref{XSbd4}) and combining Eqs. (\ref{XSEz1}),(\ref{XSFAB}), we obtain the constraints on the superposition coefficients:
\begin{equation}\label{XSbb1}
\overline{H}_{B}\left(
       \begin{array}{c}
         \overline{c}_{1} \\
         \overline{c}_{2} \\
       \end{array}
     \right)
=0
\end{equation}
with
\begin{equation}
\overline{H}_{B}=\left(
\begin{array}{cc}
(t_{1}-(\delta-\gamma)z_{1}^{N})\overline{\phi} _{A}^{(1)} & (t_{1}-(\delta-\gamma)z_{2}^{N})\overline{\phi} _{A}^{(2)} \\
\big(t_{1}z_{1}^{N}-(\delta+\gamma)\big)z_{1}\overline{\phi} _{B}^{(1)} & \big(t_{1}z_{2}^{N}-(\delta+\gamma)\big)z_{2}\overline{\phi} _{B}^{(2)}%
\end{array}%
\right).
\end{equation}
For nontrivial solutions of $(\overline{c}_1, \overline{c}_2)$, $\mathrm{det}[\overline{H}_{B}] =0$ yields the following condition:
\begin{equation}\label{XSqqz1}
\begin{split}
&\eta_1(z_{1}-z_{2})+\eta_2(z_{1}^{N-1}-z_{2}^{N-1})-\eta_3(z_{1}^{N}-z_{2}^{N})\\
=&(z_{1}^{N+1}-z_{2}^{N+1}),
\end{split}
\end{equation}
where $\eta_1= \frac{2\delta}{t_{1}}$, $\eta_2=\frac{\delta^2-\gamma^2}{t_{1}^{2}}$, $\eta_3=\frac{t_{1}^2-\delta^2+\gamma^2}{t_{1}t_{2}}$. Equation (\ref{XSqqz1}) together with  Eq. (\ref{XSz1z2}) determines the solutions of $z_1$ and $z_2$. From Eq. (\ref{XSz1z2}), we set $z_{1}=e^{i\theta}$, $z_{2}=e^{-i\theta}$. The energy spectrum Eq. (\ref{XSEz1}) then becomes
\begin{equation}
E=\pm\sqrt{2t_{1}t_{2}\cos \theta+t_{1}^2+t_{2}^2}.
\end{equation}
And Eq. (\ref{XSqqz1}) reduces to
\begin{equation}\label{XSeq-theta}
\sin[(N+1)\theta]+\eta_3\sin(N\theta)-\eta_2\sin[(N-1)\theta]-\eta_1\sin[\theta]=0.
\end{equation}
Depending on $\eta_1$,  $\eta_2$ and $\eta_3$, the solution of $\theta$ of the above equation may take real or complex values.

It is worth discussing the special case with $\overline{c}_2=0$, i.e., the eigenfunction contains only the $z_1$ solution. From the constraint Eq. (\ref{XSbb1}), we have
\begin{equation}
z_{1}^{N}=\frac{t_{1}}{(\delta-\gamma)},~~~~z_{1}^{N}=\frac{(\delta+\gamma)}{t_{1}}.
\end{equation}
This condition can only be satisfied when
\begin{equation}\label{Scdtn}
\gamma^2=\delta^2-t_{1}^2,
\end{equation}
which is equal to $\gamma=\gamma_a=\sqrt{\delta^2-t_{1}^2}$.
The solution of $z_1$ is then
\begin{equation}
z_1=e^{i\theta}=\sqrt[N]{\mu}e^{i\frac{2l\pi}{N}},
\end{equation}
where $\mu=\frac{\delta}{t_{1}}+\sqrt{(\frac{\delta}{t_1})^2-1}$, $l=0,1,2,\cdots,N-1$, and $\theta=\theta_R+i\theta_I=\frac{2l\pi}{N}-i\frac{\log{\mu}}{N}$. The energy spectrum is given by
\begin{equation}
E=\pm \sqrt{t_{1}t_{2}\bigg(\sqrt[N]{\mu}e^{i\theta_R}+\frac{1}{\sqrt[N]{\mu}}e^{-i\theta_R}\bigg)+t_{1}^2+t_{2}^2},
\end{equation}
with $\theta_R=\frac{2l\pi}{N}~(l=0,1,2,\cdots,N-1)$. All $\theta$-solutions are complex with $\theta_I=-\frac{\log{\mu}}{N}\propto\frac{1}{N}$. The eigenvalues are complex except for $\theta_R=0$. Thus, there are $2(N-1)$ complex eigenenergies and $2$ real energies. The eigenstates can be expressed as
\begin{equation}
\begin{split}
\overline{\Psi}=&[(\sqrt[N]{\mu}e^{i\theta_R})^{N-m+1}\overline{\phi} _{A}^{(i)},(\sqrt[N]{\mu}e^{i\theta_R})^{N-m+2}\overline{\phi} _{B}^{(i)},\cdots,\\
&
(\sqrt[N]{\mu}e^{i\theta_R})^{N}\overline{\phi} _{A}^{(i)},\sqrt[N]{\mu}e^{i\theta_R}\overline{\phi} _{B}^{(i)},\cdots,\\
&(\sqrt[N]{\mu}e^{i\theta_R})^{N-m}\overline{\phi} _{A}^{(i)},(\sqrt[N]{\mu}e^{i\theta_R})^{N-m+1}\overline{\phi} _{B}^{(i)}]^T.
\end{split}
\end{equation}
As $|z_1|=|\sqrt[N]{\mu}|\neq 1$, the spatial profiles of all eigenstates decay away from the impurity in a unidirectional way.

\subsection{Solutions for the case of $t_1=t_2$}
We specify the simple case with $t_1=t_2=t$ in this subsection. Without loss of generality, we set $t>0, \delta>0, \gamma>0$. For this case, $\eta_3=1-\eta_2$. The eigenvalues can be reduced to
\begin{equation}\label{XSspectrum2}
E=\pm2t\cos\big(\frac{\theta}{2}\big).
\end{equation}
Equation (\ref{XSeq-theta}) reduces to
\begin{equation}\label{XSeq-thetas}
\sin[(N+\frac{1}{2})\theta]-\eta_1\sin(\frac{\theta}{2})-\eta_2\sin[(N-\frac{1}{2})\theta]=0,
\end{equation}
where $\eta_1= \frac{2\delta}{t}$, $\eta_2=\frac{\delta^2-\gamma^2}{t^{2}}$. The solution $\theta=\theta_R+i\theta_I$ of Eq. (\ref{XSeq-thetas}) may take real or complex values depending on $\eta_1$,  and  $\eta_2$.
If $\theta\in \mathbb{R}$, we have $E\in \mathbb{R}$ and $|z_{1}|=|z_{2}|=1$, which indicates that the corresponding eigenstate is an extended state. If $\theta\in \mathbb{C}$, we have $E\in \mathbb{C}$ (except for $\theta_R=0$) and $|z_{1}|\neq 1,~|z_{2}|\neq 1$, which indicates that the corresponding eigenstate is not extended. By inserting Eq. (\ref{XSspectrum2}) into Eq. (\ref{XSFAB}), we have
\begin{eqnarray}\label{XSFABs}
\overline{\phi} _{B}^{(i)}=\pm z_i^{-1/2}\overline{\phi} _{A}^{(i)}.
\end{eqnarray}
Here the ``$\pm$" sign is consistent with ``$\pm$" in the expression of $E$. The ansatz wave function $\overline{\Psi} _{i}$ can be rewritten as
\begin{equation}
\begin{split}
\overline{\Psi} _{i}=&\big(z_{i}^{N-m+1},\pm z_{i}^{N-m+3/2}, z_{i}^{N-m+2},\pm z_{i}^{N-m+5/2},
\cdots,\\
&z_{i}^{N},\pm z_{i}^{1/2},\cdots,z_{i}^{N-m},\pm z_{i}^{N-m+1/2}\big)^T\overline{\phi} _{A}^{(i)}.
\end{split}
\end{equation}
Obviously, the wave function is extended when $|z_i|=1$. For the superimposed eigenstate described by Eq. (\ref{XSWave}), its spatial component can be rewritten as
\begin{equation}
\begin{split}
\overline{\psi}_{n,A}&=
\left\{
  \begin{array}{ll}
    \sum_{i=1}^{2}(c_{i}z_{i}^{N-m+n}), & \hbox{$1\leq n\leq m$;} \label{XSwave1}\\
    \sum_{i=1}^{2}(c_{i}z_{i}^{n-m}), & \hbox{$m< n\leq N$;}
  \end{array}
\right.\\
\overline{\psi}_{n,B}&=
\left\{
  \begin{array}{ll}
   \pm \sum_{i=1}^{2}(c_{i}z_{i}^{N-m+n+1/2}), & \hbox{$1\leq n< m$;} \\
   \pm \sum_{i=1}^{2}(c_{i}z_{i}^{n-m+1/2}), & \hbox{$m\leq n\leq N$;}
  \end{array}
\right.
\end{split}
\end{equation}
with $c_i=\overline{c}_i\overline{\phi} _{A}^{(i)}$.

In the following, we analyze the solution of $\theta$ in Eq. (\ref{XSeq-thetas}) as the non-Hermitian strength $\gamma$ varies. We set $f_1(\theta)=\sin[(N+\frac{1}{2})\theta]-\eta_2\sin[(N-\frac{1}{2})\theta]$ and $f_2(\theta)=\eta_1\sin(\frac{\theta}{2})$, and Eq. (\ref{XSeq-thetas}) reduces to
\begin{equation}\label{XSeq-thetaa}
f_1(\theta)=f_2(\theta).
\end{equation}
\begin{figure}[h]
\includegraphics[width=0.46\textwidth]{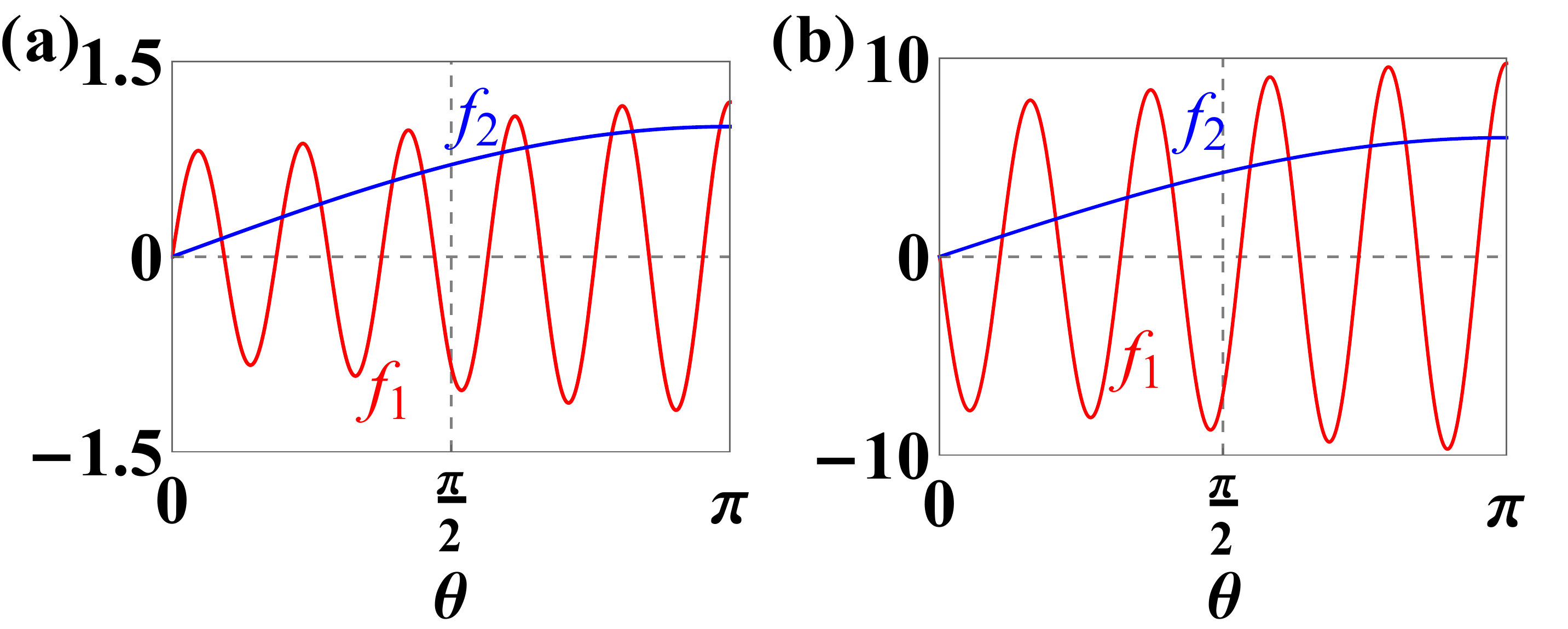}
\caption{Function $f_1(\theta)$ and $f_2({\theta})$ and their intersections. (a) $N=10,~t=1,~\delta=0.5,~\gamma=0.1$. (b) $N=10$, $t=1$, $\delta=3,$ $\gamma=1$.}
\label{fig9}
\end{figure}
The intersections of $f_1$ and $f_2$ determine the real solutions of $\theta$, as exemplified in Fig. \ref{fig9}. For small $\gamma$, there are $N$ real roots in $\theta\in(0,\pi)$ corresponding to extended bulk states when $\delta<t$, as depicted in Fig. \ref{fig9}(a), while there are at most $(N-1)$ real roots of $\theta$ in $\theta\in(0,\pi)$ when $\delta>t$, as depicted in Fig. \ref{fig9}(b). As $\gamma$ increases, the number of  intersections of $f_1$ and $f_2$, i.e., the real solutions of $\theta$, will shrink first, reach its minimum and then increase. The first disappearance and the last reemergence of the intersections occurs at $\theta=\pi$. Thus, the condition of $N$ real roots (for $\delta<t$) and $(N-1)$ real roots (for $\delta>t$) is determined by $|f_1(\theta=\pi)|>|f_2(\theta=\pi)|$, yielding $|1+\eta_2|>\eta_1$. This condition is satisfied when
\begin{equation}\label{maxroot}
\gamma<\gamma_{c_1}~~\mathrm{and}~~\gamma>\gamma_{c_2},
\end{equation}
with $\gamma_{c_1}=|\delta-t|$ and $\gamma_{c_2}=\delta+t$. As long as Eq. (\ref{maxroot}) is satisfied, there are at least $(N-1)$ real roots for Eq. (\ref{XSeq-thetaa}). Further, we discuss the region with all complex $\theta$-solutions. As $\gamma$ increases, the last disappearance and the first reemergence of the intersections occurs nearly $\theta=0$. Thus, the condition of $N$ complex roots is determined by $|f_1'(\theta=0)|<|f_2'(\theta=0)|$ with $f_i'=\frac{\partial f_i}{\partial\theta}$, giving rise to $|N(1-\eta_2)+\frac{1}{2}(1+\eta_2)|<\eta_1$. When $\eta_2=1$ (i.e., $\gamma=\gamma_a=\sqrt{\delta^2-t^2}$), the condition always is satisfied independent of $N$. In fact, this condition is satisfied in a narrow region near $\gamma=\gamma_a$ for finite $N$, and the narrow region shrinks to  $\gamma=\gamma_a$ in the thermodynamic limit.

We proceed to study the $\mathcal{PT}$-transitions of the system as $\gamma$ increase for fixed $t$ and $\delta$. There are three difference cases as listed below.

(i) \textbf{$\mathcal{PT}$-unbroken regime, $\gamma<\gamma_{c_1}$}. (1) First, if $\delta>t$ and $\gamma=0$ (Hermitian limit), Eq. (\ref{XSeq-thetaa}) has $(N-1)$ real roots corresponding to extended bulk states and a complex root with $\theta_R=0$ corresponding to a pair of bound states residing at the impurity. As expected for the Hermitian impurity, all eigenenergies are real. There are $(2N-2)$ extended bulk states except for a pair of bound states. The scenario persists even for $\gamma\neq 0$, provided that $\gamma<\gamma_{c_1}$ is satisfied. (2) Second, if $\delta<t$, as long as $\gamma<\gamma_{c_1}$, Eq. (\ref{XSeq-thetaa}) has $N$ real roots corresponding to extended bulk states (no bound state exists), and all eigenvalues are real. Combining (1)(2), all eigenvalues are real when $\gamma<\gamma_{c_1}$, the system is in the $\mathcal{PT}$-unbroken phase, with at least $2(N-1)$ extended bulk states and at most $2$ bound states.

(ii) \textbf{$\mathcal{PT}$-broken regime, $\gamma_{c_1}<\gamma<\gamma_{c_2}$}. Increasing $\gamma$ to enter this regime, the number of real roots shrinks first, reaches its minimum and then increases. The complex roots of $\theta$ give rise to complex eigenenergies and the system is in the $\mathcal{PT}$-broken phase. The number of complex eigenenergies reaches its maximum $2(N-2)$ at $\gamma=\gamma_a\equiv \sqrt{\delta^2-t^2}$. Based on the discussions in the previous section, the solutions of $z_1$ are $z_1=e^{i\theta}=\sqrt[N]{\mu}e^{i\frac{2l\pi}{N}},~~~~(l=0,1,2,\cdots,N-1)$, with $\mu=\frac{\delta}{t}+\sqrt{(\frac{\delta}{t})^2-1}$, and $\theta=\theta_R+i\theta_I=\frac{2l\pi}{N}-i\frac{\log{\mu}}{N}$. The eigenenergies are given by $E= \pm 2t\cos[\frac{2l\pi}{N}-i\log(\sqrt[N]{\mu})]$, and we have $|\mathrm{Im}(E)|=(\mu^{1/(2N)}-\mu^{-1/(2N)})|\sin\theta_R|\approx\frac{\log\mu}{N}|\sin\theta_R|$. Obviously, the local non-Hermitian term contributes a $1/N$-order correction to the imaginary part of the $\theta$-roots as well as the eigenenergies (except for $\theta_R=0$). The associated wave function $\overline{\Psi}$ for the non-Hermitian SSH model is  %and its modulo square
\begin{eqnarray}
\begin{split}
\overline{\Psi}\sim\left(
          \begin{array}{c}
             (\sqrt[N]{\mu}e^{i\theta_R})^{N-m+1} \\
             \pm (\sqrt[N]{\mu}e^{i\theta_R})^{N-m+3/2}\\
             (\sqrt[N]{\mu}e^{i\theta_R})^{N-m+2}\\
             \pm (\sqrt[N]{\mu}e^{i\theta_R})^{N-m+5/2}\\
            \vdots \\
             (\sqrt[N]{\mu}e^{i\theta_R})^{N-1}\\
             \pm (\sqrt[N]{\mu}e^{i\theta_R})^{N-1/2}\\
             (\sqrt[N]{\mu}e^{i\theta_R})^{N}\\
             \pm (\sqrt[N]{\mu}e^{i\theta_R})^{1/2}\\
             (\sqrt[N]{\mu}e^{i\theta_R})\\
             \pm (\sqrt[N]{\mu}e^{i\theta_R})^{3/2}\\
            \vdots \\
             (\sqrt[N]{\mu}e^{i\theta_R})^{N-m}\\
            \pm (\sqrt[N]{\mu}e^{i\theta_R})^{N-m+1/2}
          \end{array}
        \right).\\
\end{split}
\end{eqnarray}
%\begin{eqnarray}
%\begin{split}
%|\overline{\Psi}|^2\sim\left(
%          \begin{array}{c}
%             (\sqrt[N]{\mu})^{2(N-m)+2} \\
%             (\sqrt[N]{\mu})^{2(N-m)+3}\\
%             (\sqrt[N]{\mu})^{2(N-m)+4}\\
%             (\sqrt[N]{\mu})^{2(N-m)+5}\\
%            \vdots \\
%             (\sqrt[N]{\mu})^{2N-2}\\
%             (\sqrt[N]{\mu})^{2N-1}\\
%             (\sqrt[N]{\mu})^{2N}\\
%             \sqrt[N]{\mu}\\
%             (\sqrt[N]{\mu})^{2}\\
%             (\sqrt[N]{\mu})^{3}\\
%            \vdots \\
%             (\sqrt[N]{\mu})^{2(N-m)}\\
%             (\sqrt[N]{\mu})^{2(N-m)+1}
%          \end{array}
%        \right).
%\end{split}
%\end{eqnarray}
Thus  all wave functions have the same spatial profiles:
\begin{equation}
\begin{split}
|\overline{\psi}_{x}|&=
\left\{
  \begin{array}{ll}
    \mu^{\frac{x-x_{mA}}{2N}+1}, & \hbox{$x\leq x_{mA}$;}\\
    \mu^{\frac{x-x_{mA}}{2N}}, & \hbox{$x> x_{mA}$;}
  \end{array}
\right.
\end{split}
\end{equation}
where $x=1,\cdots,2N$, and $x_{mA}=2m-1$. Denote $\xi$ as the localization length of the eigenstate: $|\overline{\psi}_{x}|\sim e^{\frac{x-x_{mA}}{\xi}}$. It is easy to see $\xi=\frac{2N}{\log \mu}$, which is proportional to the system size. These eigenstates are dubbed scale-free localized (SFL) states in the main text, which decay away from the impurity in a unidirectional way. They differ from the usual non-Hermitian skin modes that have finite localization length even when $N\rightarrow\infty$. The eigenstates for the double chain model can be easily obtained by the transformation $|\Psi\rangle=S^{-1}|\overline{\Psi}\rangle$ as
\begin{eqnarray}
\begin{split}
\Psi\sim\left(
          \begin{array}{c}
            (\sqrt[N]{\mu}e^{i\theta_R})^{N-m+1}\mp i(\sqrt[N]{\mu}e^{i\theta_R})^{N-m+3/2} \\
            -i(\sqrt[N]{\mu}e^{i\theta_R})^{N-m+1}\pm (\sqrt[N]{\mu}e^{i\theta_R})^{N-m+3/2} \\
            \vdots \\
            (\sqrt[N]{\mu}e^{i\theta_R})^{N-1}\mp i(\sqrt[N]{\mu}e^{i\theta_R})^{N-1/2}\\
            -i(\sqrt[N]{\mu}e^{i\theta_R})^{N-1}\pm (\sqrt[N]{\mu}e^{i\theta_R})^{N-1/2} \\
            (\sqrt[N]{\mu}e^{i\theta_R})^{N}\mp i(\sqrt[N]{\mu}e^{i\theta_R})^{1/2}\\
            -i(\sqrt[N]{\mu}e^{i\theta_R})^{N}\pm (\sqrt[N]{\mu}e^{i\theta_R})^{1/2} \\
            (\sqrt[N]{\mu}e^{i\theta_R})\mp i(\sqrt[N]{\mu}e^{i\theta_R})^{3/2}\\
            -i(\sqrt[N]{\mu}e^{i\theta_R})\pm (\sqrt[N]{\mu}e^{i\theta_R})^{3/2} \\
            \vdots \\
            (\sqrt[N]{\mu}e^{i\theta_R})^{N-m}\mp i(\sqrt[N]{\mu}e^{i\theta_R})^{N-m+1/2}\\
            -i(\sqrt[N]{\mu}e^{i\theta_R})^{N-m}\pm (\sqrt[N]{\mu}e^{i\theta_R})^{N-m+1/2}
          \end{array}
        \right).~~~~
\end{split}
\end{eqnarray}
%\begin{eqnarray}
%\begin{split}
%|\Psi|^2\sim\left(
%          \begin{array}{c}
%            (\sqrt[N]{\mu})^{2(N-m+1)}[1+\sqrt[N]{\mu}\pm 2\sqrt[2N]{\mu}\sin(\frac{\theta_R}{2})] \\
%            (\sqrt[N]{\mu})^{2(N-m+1)}[1+\sqrt[N]{\mu}\mp 2\sqrt[2N]{\mu}\sin(\frac{\theta_R}{2})] \\
%            \vdots \\
%            (\sqrt[N]{\mu})^{2(N-1)}[1+\sqrt[N]{\mu}\pm 2\sqrt[2N]{\mu}\sin(\frac{\theta_R}{2})]\\
%            (\sqrt[N]{\mu})^{2(N-1)}[1+\sqrt[N]{\mu}\mp 2\sqrt[2N]{\mu}\sin(\frac{\theta_R}{2})] \\
%            \mu^2+\sqrt[N]{\mu}\pm2\mu\sqrt[2N]{\mu}\sin[(1/2-N)\theta_R]\\
%            \mu^2+\sqrt[N]{\mu}\mp2\mu\sqrt[2N]{\mu}\sin[(1/2-N)\theta_R]\\
%            (\sqrt[N]{\mu})^2[1+\sqrt[N]{\mu}\pm 2\sqrt[2N]{\mu}\sin(\frac{\theta_R}{2})]\\
%            (\sqrt[N]{\mu})^2[1+\sqrt[N]{\mu}\mp 2\sqrt[2N]{\mu}\sin(\frac{\theta_R}{2})] \\
%            \vdots \\
%            (\sqrt[N]{\mu})^{2(N-m)}[1+\sqrt[N]{\mu}\pm 2\sqrt[2N]{\mu}\sin(\frac{\theta_R}{2})] \\
%            (\sqrt[N]{\mu})^{2(N-m)}[1+\sqrt[N]{\mu}\mp 2\sqrt[2N]{\mu}\sin(\frac{\theta_R}{2})]
%          \end{array}
%        \right).
%\end{split}
%\end{eqnarray}

To conclude, in the $\mathcal{PT}$-broken regime, the number of complex eigenenergies ranges from $2$ to $2(N-1)$, with their associated eigenstates being SFL states. In particular, when $\gamma=\gamma_a$ (in fact, in a narrow region near $\gamma_a$ for finite $N$), the number of complex eigenenergies reaches its maximum $2(N-1)$ and all eigenstates are SFL states.

(iii) \textbf{$\mathcal{PT}$-restoration regime, $\gamma>\gamma_{c_2}$}. In this regime, the number of real roots of Eq. (\ref{XSeq-thetaa}) recovers to $(N-1)$. Besides, there is a complex root with $\theta_{R}=\pi$ corresponding to a pair of bound states with complex eigenenergies. Therefore, there are $2(N-1)$ real eigenenergies corresponding to extended bulk states and $2$ bound states with complex eigenenergies.
\begin{figure*}[b]
\includegraphics[width=.95\textwidth]{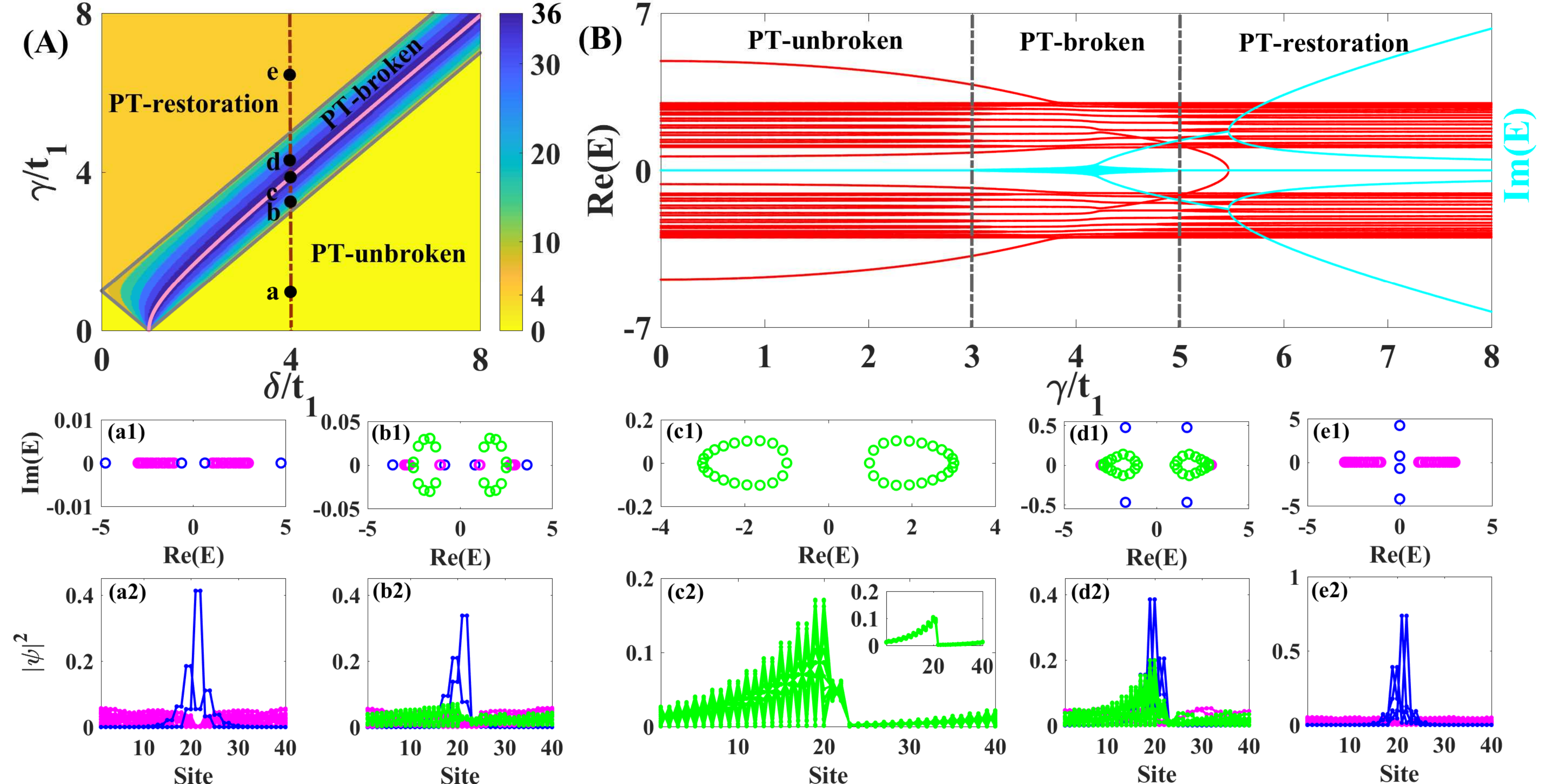}
\caption{(A) Phase diagram for the double chain model. Boundaries of different regimes are marked by gray lines. In different phase regimes, The color-coded numbers represent the number of complex eigenenergies for a finite-size lattice with $2N=40$. Along the pink line $\gamma=\gamma_a=\sqrt{\delta^2-t_1^2}$, all eigenstates are SFL states. (B) Energy spectra for the double chain model versus $\gamma$ with fixed $\delta/t_1=4$ [See the brown line in (A)]. Red/cyan lines represent real/imaginary parts of eigenenergies. (a1-e1) Energy spectra on the complex-energy plane with $\gamma=1,~3.2,~\sqrt{15},~4.3,~6.5$ corresponding to dots `a-e' in (A), respectively. (a2-e2) The associated spatial profiles of all eigenstates. The inset in (c2) plots the spatial profiles of eigenstates for the non-Hermitian SSH model with the same parameters. In (a1-e1, a2-e2), the blue/magenta/green data represent bound states/extended states/SFL states, respectively. Other parameters are $2N=40$, $t_1=1$, $t_2=2$, and the impurity rung is set at $m=N/2+1$.}
\label{fig10}
\end{figure*}
\begin{figure*}[hbt]
\includegraphics[width=.98\textwidth]{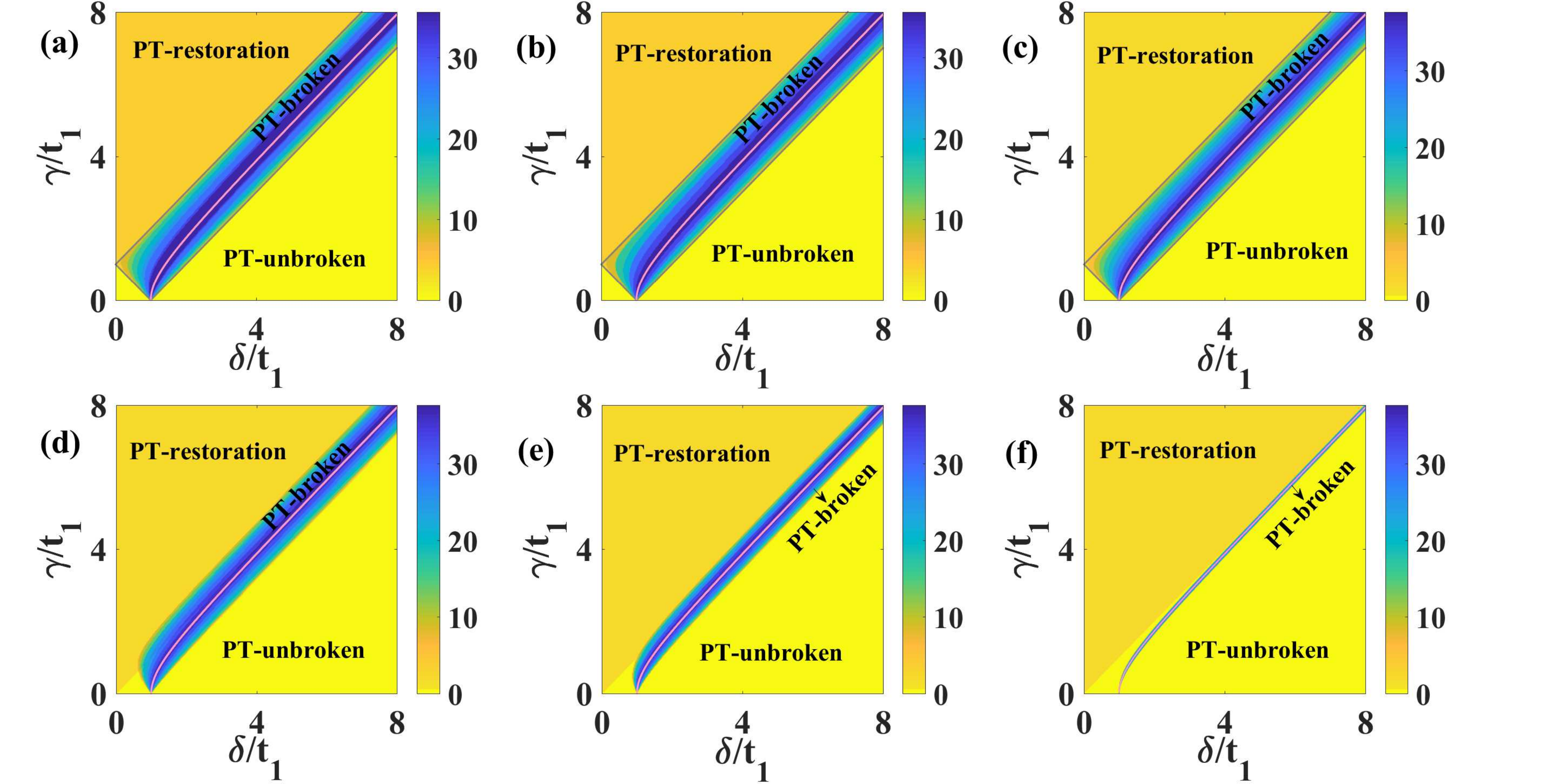}
\caption{Phase diagram for the double chain model with different $t_2$. Along the pink line $\gamma=\gamma_a=\sqrt{\delta^2-t_1^2}$, all eigenstates are SFL states, whose spatial profiles decay away from the impurity in a unidirectional way. (a) $t_2=10$; (b) $t_2=2$; (c) $t_2=1$; (d) $t_2=0.8$; (e) $t_2=0.5$; (f) $t_2=0.1$. The common parameters: $2N=40$, $t_1=1$, and the impurity rung is set at $m=N/2+1$.}
\label{fig11}
\end{figure*}

\section{Emergence of SFL states of the double chain model for the $t_1 \neq t_2$ case}\label{Appendix C}
In the main text, we have demonstrated the phase diagram [See Fig. \ref{fig1}(a)] and $\mathcal{PT}$-symmetry breaking for $t_1=t_2$. Here we turn to the generic case with $t_1 \neq t_2$ and show that the $\mathcal{PT}$-symmetry breaking and the emergence of SFL states also occur. In Fig. \ref{fig10}, we display the phase diagram for the double chain model with $t_1=1,~t_2=2$. There also exist three distinct regimes, i.e., $\mathcal{PT}$-unbroken, $\mathcal{PT}$-broken, and $\mathcal{PT}$-restoration with a little subtlety. Their boundaries are determined by $\gamma_{c_1}=|\delta-t_1|,~\gamma_{c_2}=\delta+t_1$, as marked in gray lines in Fig. \ref{fig10}(A). Figure \ref{fig10}(B) plots the spectrum versus $\gamma$ by fixing $\delta/t_1=4$. In the $\mathcal{PT}$-unbroken regime with $\gamma<\gamma_{c_1}$, there are $2(N-2)$ extended bulk states corresponding to $N-2$ real roots of Eq. (\ref{XSeq-theta}) and two pairs of bound states located at impurity corresponding to $2$ pure imaginary roots. All eigenenergies are real as shown in Figs. \ref{fig10}(a1),(a2). The system is in $\mathcal{PT}$-broken phase when $\gamma_{c_1}<\gamma<\gamma_{c_2}$, where $\theta$ has complex roots corresponding to complex eigenenergies. In this regime, the number of real roots of $\theta$ decreases first, reaches its minimum at $\gamma=\gamma_a=\sqrt{\delta^2-t_1^2}$ and then increases with increasing $\gamma$. At $\gamma=\gamma_a$, the eigenfunction contains only the $z_1$ solution due to $\overline{c}_2=0$, and $|z_1|=|\sqrt[N]{\mu}|=\sqrt[N]{\frac{\delta}{t_1}+\sqrt{(\frac{\delta}{t_1})^2-1}}> 1$. Thus all eigenstates are SFL states, decaying away from the impurity in a unidirectional way as depicted in Figs. \ref{fig10}(c1),(c2). For $\gamma\neq\gamma_a$ in the $\mathcal{PT}$-broken regime, the extended state with real eigenenergies and SFL states with complex eigenenergies coexist as shown in Figs. \ref{fig10}(b1),(b2),(d1),(d2). In short, the number of complex eigenenergies $N_{\mathrm{Im}}=4\sim2(N-2)$, and there are $(4\sim 2N)$ SFL states in the $\mathcal{PT}$-broken regime. In the $\mathcal{PT}$-restoration reigme with $\gamma>\gamma_{c_2}$, the number of real roots of Eq. (\ref{XSeq-theta}) recovers to $(N-2)$. In addition, there are $2$ complex roots of Eq. (\ref{XSeq-theta}). Thus, there are $2(N-2)$ extended bulk states with real eigenenergies and $4$ bound states at the impurity with complex eigenenergies, as shown in Figs. \ref{fig10}(e1),(e2).

In Fig. \ref{fig11}, we display phase diagram for the double chain model with different $t_2$. Except of the extreme case $t_2=0$, there always exists a $\mathcal{PT}$-broken regime surrounding $\gamma=\gamma_a$, accompanied by the emergence of SFL states, for various $t_2$ as displayed in Fig. \ref{fig11}. In particular, when $\gamma=\gamma_a$, all eigenstates are SFL states, whose spatial profiles decay away from the impurity in a unidirectional way. Explicitly, when $t_2\geq t_1$, the phase boundary are determined by $\gamma_{c_1}=|\delta-t_1|,~\gamma_{c_2}=\delta+t_1$. When $t_2<t_1$, the $\mathcal{PT}$-broken region with SFL states gradually decrease as $t_2$ decreases, and this region get closer and closer around $\gamma=\gamma_a$. Imaging the extreme situation with $t_2=0$, there is no SFL state as expected because the part of local non-Hermitian is not connected to other parts of bulk. In addition, the emergence of SFL states requires that $\delta$ can not be too small when $t_2 < t_1$. Our results reveal that local non-Hermiticity generated scale-free localization is a general phenomenon even for multi-band systems.

\section{Exact solutions of the single-impurity model}\label{Appendix D}
The Hamiltonian of the single-impurity model is given by Eq. (\ref{impurity}).
The corresponding eigenproblem reads
\begin{equation}\label{eigen_h0}
\hat{H}|\Psi\rangle=E|\Psi\rangle,
\end{equation}
with $|\Psi\rangle=\sum\limits_{n=1}^{L} (\psi_{n} \hat{c}_{n}^{\dag } ) |0\rangle$. In the form of its component, $\Psi=(\psi_{1},\psi_{2},\cdots,\psi_{m},\cdots,\psi_{L-1},\psi_{L})^T$. Equation (\ref{eigen_h0}) consists of a series of bulk equations and the impurity equations. The bulk equations are given by
\begin{equation}
t\psi_{s-1}-E\psi_{s}+t\psi_{s+1}=0\label{XSbk1AA}\\
\end{equation}
with $s=1,\cdots,m-1,m+2,\cdots,L$.
The impurity equations are given by
\begin{eqnarray}
t\psi_{m-1}-(E-i\gamma)\psi_{m}+t\psi_{m+1} &=&0,\label{XSbd3AA}\\
t\psi_{m}-E\psi_{m+1}+t\psi_{m+2} &=&0.\label{XSbd4AA}
\end{eqnarray}
We take the ansatz wave function $\Psi _{i}$ satisfying the bulk Eqs. (\ref{XSbk1AA}) as follows
\begin{equation}\label{XSFiiAA}
\begin{split}
\Psi_{i}=&(z_i^{L-m+1},z_i^{L-m+2},\cdots,z_i^{L-1},z_i^{L},z_i,\\
&\cdots,z_i^{L-m-1},z_i^{L-m})^T.
\end{split}
\end{equation}
Inserting Eq. (\ref{XSFiiAA}) into the Eq. (\ref{XSbk1AA}) yields the expression of eigenvalue in terms of $z_i$:
\begin{equation}\label{XSEz1AA}
E = t\bigg(z+\frac{1}{z}\bigg).
\end{equation}
For a given $E$, there are two solutions of $z_i$ (denoted as $z_1, z_2$) and they fulfill the constraint:
\begin{eqnarray}\label{XSz1z2AA}
z_{1}z_{2} &=&1.
\end{eqnarray}
The wave function should be the superposition:
\begin{equation}
\begin{split}
\Psi=c_{1}\Psi _{1}+c_{2}\Psi_{2}=(\psi_{1},\psi_{2},\cdots,\psi_{m},\cdots,\psi_{L-1},\psi_{L})^T,~~~ \label{XSWaveAA}
\end{split}
\end{equation}
where
\begin{equation}
\begin{split}
\psi_{n}=
\left\{
  \begin{array}{ll}
    \sum_{i=1}^{2}(c_{i}z_{i}^{L-m+n}),  \hbox{$1\leq n\leq j$;} \label{XSwave1AA}\\
    \sum_{i=1}^{2}(c_{i}z_{i}^{n-m}), ~~~ \hbox{$j< n\leq L$.}
  \end{array}
\right.
\end{split}
\end{equation}
By inserting it into Eqs. (\ref{XSbd3AA})(\ref{XSbd4AA}) and combining Eq. (\ref{XSEz1AA}), the impurity equation transforms into
\begin{equation}\label{XSbb1AA}
H_{B}\left(
       \begin{array}{c}
         c_{1} \\
         c_{2} \\
       \end{array}
     \right)=0
\end{equation}
with
\begin{equation}\label{XSbb1AA}
H_{B}=\left(
\begin{array}{cc}
1-z_{1}^{L} & 1-z_{2}^{L} \\
tz_{1}(1-z_1^L)+i\gamma z_1^L & tz_{2}(1-z_2^L)+i\gamma z_2^L
\end{array}%
\right).
\end{equation}
The nontrivial solutions of $(c_1, c_2)$ are determined by $\mathrm{det}[H_{B}] =0$, yielding
\begin{equation}\label{XSqqz1AA}
\begin{split}
t(2-z_1^L-z_2^L)(z_1-z_2)+i\gamma(z_1^L-z_2^L)=0.
\end{split}
\end{equation}
Equation (\ref{XSqqz1AA}) and  Eq. (\ref{XSz1z2AA}) together determine the solutions of $z_1$ and $z_2$. From Eq. (\ref{XSz1z2AA}), we set $z_{1}=e^{i\theta}$, $z_{2}=e^{-i\theta}$, then Eq. (\ref{XSqqz1AA}) becomes
\begin{equation}\label{XSeq-thetaAA}
\sin\big(\frac{L\theta}{2}\big)\bigg[2t\sin\theta\sin\big(\frac{L\theta}{2}\big)+i\gamma\cos\big(\frac{L}{2}\theta\big)\bigg]=0.
\end{equation}
And the corresponding eigenenergies are expressed as
\begin{equation}\label{XSspectrum1AA}
E=2t\cos \theta.
\end{equation}
By observing Eq. (\ref{XSeq-thetaAA}), we have two types of solutions. The first type is from $\sin\big(\frac{L\theta}{2}\big)=0$. The roots are $\theta=\frac{2l\pi}{L}$ with $l=1,2,\cdots,L/2-1$ for even $L$, and $l=1,2,\cdots,(L-1)/2$ for odd $L$. Thus there are $L/2-1$ real eigenenergies for $L\in \mathrm{even}$, and $(L-1)/2$ real eigenenergies for $L\in \mathrm{odd}$. Their corresponding eigenstates are all extended and
unaffected by the local on-site imaginary potential.

The other eigenstates come from the second type of solutions:
\begin{equation}\label{XSspectrum1AA2}
2t\sin\theta\sin\big(\frac{L\theta}{2}\big)+i\gamma\cos\big(\frac{L}{2}\theta\big)=0.
\end{equation}
The solutions $\theta$ of above equation has complex roots, denoted as $\theta=\theta_R+i\theta_I$. The number of complex eigenenergies are $L/2+1$ for even $L$ and $(L+1)/2$ for odd $L$. To obtain them, let us first assume $\theta_I\propto L^0$, then we have
\begin{equation}\label{appAA}
\begin{split}
&\sin\big(\frac{L\theta}{2}\big)\approx\frac{i}{2}\mathrm{sgn}(\theta_I)e^{-\frac{i}{2}\mathrm{sgn}(\theta_I)L\theta},\\
&\cos\big(\frac{L\theta}{2}\big)\approx\frac{1}{2}e^{-\frac{i}{2}\mathrm{sgn}(\theta_I)L\theta}
\end{split}
\end{equation}
for large $L$. Inserting Eq. (\ref{appAA}) into  Eq. (\ref{XSspectrum1AA2}), we get $2t\sin\theta\mathrm{sgn}(\theta_I)+\gamma=0$. It has a root with $\theta_I<0$ only if $\frac{\gamma}{2t}>1$. Explicitly, the solution is written as
\begin{equation}
\theta=\frac{\pi}{2}+i\mathrm{arcosh}\big(\frac{\gamma}{2t}\big),
\end{equation}
which satisfies $\theta_I\propto L^0$. This solution is associated with a bound state.

Let us then assume $\theta_I\propto L^{-1}$, we have $\sin\theta\approx\sin\theta_R$ for large $L$. The real and imaginary parts of Eq. (\ref{XSspectrum1AA2}) become
\begin{equation}
\begin{split}
\sin\big(\frac{L\theta_R}{2}\big)\cosh\big(\frac{L\theta_I}{2}\big)\bigg[ 2t\sin\theta_R+\gamma\tanh\big(\frac{L\theta_I}{2}\big) \bigg]=0,\\
\cos\big(\frac{L\theta_R}{2}\big)\cosh\big(\frac{L\theta_I}{2}\big)\bigg[ 2t\sin\theta_R\tanh\big(\frac{L\theta_I}{2}\big)+\gamma \bigg]=0.
\end{split}
\end{equation}
Hence we have either
\begin{equation}\label{tIAA1}
\cos\big(\frac{L\theta_R}{2}\big)=0,~~~\tanh\big(\frac{L\theta_I}{2}\big)=-\frac{2t}{\gamma}\sin\theta_R,
\end{equation}
or
\begin{equation}\label{tIAA2}
\sin\big(\frac{L\theta_R}{2}\big)=0,~~~\tanh\big(\frac{L\theta_I}{2}\big)=-\frac{\gamma}{2t}(\sin\theta_R)^{-1}.
\end{equation}
The solutions of Eqs. (\ref{tIAA1})(\ref{tIAA2}) are respectively
\begin{equation}\label{tIAA21}
\begin{split}
\theta=\frac{(2l+1)\pi}{L}+i\frac{2}{L}\mathrm{artanh}\bigg[-\frac{2t}{\gamma}\sin\big[\frac{(2l+1)\pi}{L}\big]\bigg];
\end{split}
\end{equation}
\begin{equation}\label{tIAA22}
\begin{split}
\theta&=\frac{2l\pi}{L}+i\frac{2}{L}\mathrm{artanh}\bigg[-\frac{\gamma}{2t}\big[\sin(\frac{2l\pi}{L})\big]^{-1}\bigg].
\end{split}
\end{equation}
It is clear that the imaginary part of the complex roots satisfy $\theta_I\propto\frac{1}{L}$. In another word, the localization length of these eigenstates (except for bound states) with complex roots is proportional to the system size $\xi\propto L$, and these eigenstates are SFL states.


\begin{thebibliography}{99}
\bibitem{nn} For $\overline{\psi}_{n,A}$, we have $\tilde{n}=m-n$ if $1\leq n\leq m$ and $\tilde{n}=m+N-n$ if $m<n\leq N$. For $\overline{\psi}_{n,B}$, we set $\tilde{n}=m-n$ if $1\leq n< m$ and $\tilde{n}=m+N-n$ if $m\leq n\leq N$.
\end{thebibliography}
\end{document}